\documentclass[a4paper,11pt]{article}
\usepackage[utf8]{inputenc}
\usepackage[english]{babel}
\usepackage{braket}
\usepackage{xspace}
\usepackage{bbm}
\usepackage{mathdots}
\usepackage{stackrel}
\usepackage{xcolor}
\usepackage{mathtools}
\usepackage{bbm}
\usepackage{subcaption}
\usepackage[T1]{fontenc}               
\usepackage[left=3cm,
            right=2.5cm,
            top=2.5cm,
            bottom=3cm
            ]{geometry}                
\usepackage{graphicx}
\usepackage{mathrsfs}
\usepackage{appendix}
\usepackage{fancyhdr}
\usepackage{amsmath,                   
            amssymb,                   
            amsthm}                    

\usepackage{subcaption,braket}

\usepackage[colorlinks=true,citecolor=magenta,urlcolor=magenta,linkcolor=magenta]{hyperref}

\usepackage[style=phys,hyperref=true,articletitle=true,biblabel=brackets,chaptertitle=false]{biblatex}
\usepackage{comment}

\usepackage{setspace}

\numberwithin{equation}{section}

\def \be {\begin{equation}} 
\def \ee {\end{equation}} 
\def \l {\left(} 
\def \r {\right)}

\newcommand{\TT}{{\cal T}}
\newcommand{\Tr}{\text{Tr}}
\def \ben {\begin{equation*}} 
\def \een {\end{equation*}} 

\date{}

\title{Entanglement asymmetry in the ordered phase of many-body systems: the Ising Field Theory}

\author{Luca Capizzi$^1$, Michele Mazzoni $^2$}
\begin{document}
\maketitle
$^1$SISSA and INFN Sezione di Trieste, via Bonomea 265, 34136 Trieste, Italy.\\
$^{2}$ Department of Mathematics, City, University of London, 10 Northampton Square EC1V 0HB, UK.\\
\begin{abstract} 
Global symmetries of quantum many-body systems can be spontaneously broken. Whenever this mechanism happens, the ground state is degenerate and one encounters an ordered phase. In this study, our objective is to investigate this phenomenon by examining the entanglement asymmetry of a specific region. This quantity, which has recently been introduced in the context of $U(1)$ symmetry breaking, is extended to encompass arbitrary finite groups $G$. We also establish a field theoretic framework in the replica theory using twist operators.
We explicitly demonstrate our construction in the ordered phase of the Ising field theory in 1+1 dimensions, where a $\mathbb{Z}_2$ symmetry is spontaneously broken, and we employ a form factor bootstrap approach to characterise a family of composite twist fields. Analytical predictions are provided for the entanglement asymmetry of an interval in the Ising model as the length of the interval becomes large. We also propose a general conjecture relating the entanglement asymmetry and the number of degenerate vacua, expected to be valid for a large class of states, and we prove it explicitly in some cases.
\end{abstract}

\section{Introduction}

Symmetry is nowadays considered a cornerstone of modern Physics. Its breaking is responsible for a plethora of interesting phenomena, such as ferromagnetism \cite{Kittel-18}, superconduction, superfluidity \cite{Annett-04}. Spontaneous symmetry breaking is the phenomenon by which a symmetry possessed by a quantum system, say described in terms of its Hamiltonian/Lagrangian or its equations of motion, is not mirrored by some of the corresponding states. For instance, it is very well known that, at low-enough temperatures, some ferromagnetic materials magnetise spontaneously in a given direction, which depends solely on the way those materials are cooled down. A similar mechanism has been observed at zero temperature in  \lq\lq frustrated'' quantum systems, e.g. quantum spin chains, where spontaneous symmetry breaking arises if an external parameter, such as a magnetic field or a chemical potential, is varied.

Zero-temperature phases in which the symmetry is absent or present can be separated by a quantum phase transition \cite{Sachdev-99}. Close to the transition, quantum correlations are dominant, giving rise to a large amount of entanglement among the regions of the systems. 
Much attention has been devoted in the last decade regarding the relation between symmetries and entanglement \cite{gs-18,xas-18,lr-14}, especially in the context of zero-temperature states close to phase transitions, which can be investigated quantitatively via Quantum Field Theory (QFT) \cite{mdc-20b,
crc-20,znm-21,
wznm-21}. In particular, a notion of symmetry-resolved entanglement was introduced in \cite{gs-18,xas-18,lr-14} and further developed in \cite{cgs-18,Chen-21,Chen-22,cc-21,Parez-22}. 

Little research focused on exploring the relationship between symmetry breaking and entanglement. Recently, a new measure of entanglement, dubbed \textit{entanglement asymmetry} \cite{amc-23}, was introduced to probe symmetry breaking in many-body systems. In particular, such quantity was used to analyse the restoration (or lack thereof) of a $U(1)$ symmetry in the quench dynamics of quantum spin chains \cite{amc-23,amvc-23,bkccr-23}. The purpose of this work is to apply the proposed approach of \cite{amc-23} to characterise the spontaneous symmetry-breaking pattern in quantum many-body systems at equilibrium.

We initiate this program by giving a definition of entanglement asymmetry that can be applied to any finite or compact Lie group, extending the construction already provided for $U(1)$ in \cite{amc-23} and for $\mathbb{Z}_N$ in \cite{fac-23}. Moreover, we provide a field-theoretic treatment of the one-dimensional quantum Ising model via form factor bootstrap. Our approach combines the expression of the R\'enyi entropies in terms of twist fields via the replica trick \cite{cc-09,ccd-08} and its extension in the presence of additional Aharonov–Bohm fluxes \cite{gs-18}, which stem from the action of the group and give rise to composite (charged) twist fields \cite{Levi-12}. A vast literature regarding Integrable Field Theories where similar fields were considered is present, and we refer the reader to \cite{hc-20,hcc-21,chcc-22,ccdms-22,cdmsc-22,cmc-23,cm-23,hcc-22} for further details. However, most of these works refer to paramagnetic phases of field theories, where a single symmetric vacuum is present: there, different ways of inserting the same total Aharonov–Bohm flux among the replicas give rise to the same results (see e.g \cite{hc-20,hcc-21,chcc-22}). Indeed, it has not been appreciated so far that those distinct choices correspond to distinct operators, and symmetry-broken states spot those differences explicitly. The mechanism above, which is new to our knowledge, lies at the core of entanglement asymmetry, as we will show. 

Before entering the core of this work, we first provide a definition of R\'enyi entanglement asymmetry, inspired by \cite{amc-23}, which applies to any finite group $G$. Let us consider a (possibly mixed) state $\rho$ of a bipartite system $A\cup \bar{A}$, described by the Hilbert space \cite{afov-09} 
\be
\label{Hilbert_space_bipartition}
\mathcal{H} = \mathcal{H}_A \otimes \mathcal{H}_{\bar{A}}\,.
\ee
We assume that a finite group $G$ acts unitarily on $\mathcal{H}$ via a linear map $G \ni g \mapsto \hat{g} \in \text{End}(\mathcal{H})$ satisfying 
\be
\label{G_unitary_rep}
\hat{g}= \hat{g}_A \otimes \hat{g}_{\bar{A}} \in \text{End}\l \mathcal{H}_A\r \otimes \text{End}\l \mathcal{H}_{\bar{A}}\r.
\ee
That is, $A$ and $\bar{A}$ are not mixed by $G$, which plays the role of a global symmetry for the system (see also Ref. \cite{cdm-21}). Given $\rho$, we construct the reduced density matrix over $A$ as \cite{afov-09}
\be
\rho_A \equiv \text{Tr}_{\bar{A}}\l \rho \r,
\ee
and we aim to understand whether $\rho_A$ is symmetric under the group $G$. This is tantamount to asking whether the equality
\be
\rho_A = \hat{g}_A \rho_A \hat{g}^{-1}_A
\ee
always holds or it is violated for some $g \in G$, thus signaling a breaking of the symmetry. To do so, we introduce a fictitious density matrix $\tilde{\rho}_A$ defined as
\be\label{eq:rho_tilde}
\tilde{\rho}_A \equiv \frac{1}{|G|}\sum_{g \in G} \hat{g}_A\rho_A\hat{g}^{-1}_A\,.
\ee
It is easy to show that $ \tilde{\rho}_A$ is symmetric under $G$, namely
\be
\hat{g}_A\tilde{\rho}_A\hat{g}^{-1}_A = \tilde{\rho}_A\,, \quad  \forall g \in G\,.
\ee
Moreover $\tilde{\rho}_A$ can be regarded as the \textit{symmetrisation} of $\rho_A$ under the adjoint action of the group  (see e.g. a standard textbook in linear representations of finite group \cite{Vinberg-89}).

It is not difficult to check
\be
\tilde{\rho}_A = \rho_A \quad \text{iff } \quad \rho_A = \hat{g}_A \rho_A \hat{g}^{-1}_A\,, \quad \forall g \in G\,.
\ee
Therefore, following the logic of \cite{amc-23}, it is rather natural to compare the two states $\rho_A$ and $\tilde{\rho}_A$ to probe (spontaneous or explicit) symmetry-breaking at the level of the subsystem $A$. We do so via the introduction of the \textit{R\'enyi entanglement asymmetry}, defined as
\be\label{eq:Ent_asymm}
\Delta S_n \equiv \frac{1}{1-n}\log\text{Tr}\l \tilde{\rho}^n_A\r - \frac{1}{1-n}\log\text{Tr}\l \rho^n_A\r,
\ee
that is the difference of R\'enyi entropies of the two states. Similarly, in the limit $n\rightarrow 1$ we get the difference of von Neumann entropy
\be
\Delta S_1 \equiv -\text{Tr}\l \tilde{\rho}_A \log \tilde{\rho}_A \r + \text{Tr}\l \rho_A \log\rho_A \r,
\ee
and we refer to $\Delta S_1$ as the \textit{entanglement asymmetry}. We mention that the definitions above can be straightforwardly generalised from finite groups to compact Lie groups. For instance,
given the (normalised) Haar measure $\int_{G} \mathrm{d}g$ of a compact Lie Group $G$ \cite{Vinberg-89}, it is sufficient to replace \eqref{eq:rho_tilde} with
\be
\tilde{\rho}_A \equiv \int_G \mathrm{d}g \,  \hat{g}_A\rho_A\hat{g}^{-1}_A\,,
\ee
that is compatible with the original formulation of asymmetry valid for $U(1)$ in Ref. \cite{amc-23}. While most of the theory we discuss in this work is unchanged for continuous groups, finite groups are the only relevant ones in the context of spontaneous symmetry breaking in zero-temperature one-dimensional systems, due to the Mermin-Wagner no-go theorem\cite{mw-66}.

We analyse in detail the Ising field theory \cite{mw-73,Delfino-98} in its ferromagnetic phase, probing the symmetry breaking pattern
\be
\mathbb{Z}_2 \rightarrow \{1\}\,,
\ee
via the (R\'enyi) entanglement asymmetry associated with the group $G =\mathbb{Z}_2$. In particular, we consider one (of the two) spontaneously broken ground-states, and we compute the asymmetry of an interval of size $\ell$ which is large compared to the correlation length $\sim m^{-1}$ of the model. We do so with a replica trick, relating the R\'enyi entanglement asymmetry for $n\geq 2$ integer to the expectation values of some $\mathbb{Z}_2$ composite twist fields. We describe systematically the form factors of these fields, and we get analytical results for their correlation functions in the two-particle approximation.
We briefly summarise the main result for the Ising model, that is
\be\label{eq:Asym_Ising}
\Delta S_n \simeq \log 2\,, \quad m\ell \gg 1\,,
\ee
up to exponentially small corrections that are derived explicitely.

We structure our manuscript as follows. In Sec. \ref{sec:twist_op} we provide an explicit construction of the (composite) twist operators, valid for finite-dimensional Hilbert spaces, and we relate their expectation values to the asymmetry of a subsystem. In Sec. \ref{sec:FF_Ising}, we review the scattering properties of the Ising field theory, and we characterise the form factors of the standard twist field in the ferromagnetic phase, which are eventually employed in the computation of the R\'enyi entropy. Then, in Sec. \ref{sec:FF_comp_Ising}, the core of our work, we extend our analysis to a family of composite $\mathbb{Z}_2$ twist fields. In particular, we focus on those fields with vanishing net $\mathbb{Z}_2$ flux across the replicas, and we establish a connection between their form factors and the ones of the standard twist fields. This analysis allows us to compute the entanglement asymmetry of a large interval. Finally, in Sec. \ref{sec:Further_gen} we identify a fundamental mechanism behind the large volume behavior of the entanglement asymmetry of arbitrary (clustering) states, and we provide a general conjecture valid for any finite group $G$. We leave conclusions and outlook to Sec. \ref{sec:conclusions}.


\section{Twist operators and Entanglement asymmetry}\label{sec:twist_op}

Here, we follow the idea of Refs. \cite{cd-11,cd-12} and we introduce a family of operators in the replica theory, the twist operators, which allow us to express the entanglement measures of interest as expectation values. We give a rigorous characterisation of those operators for finite-dimensional Hilbert spaces, extending the analysis of \cite{cd-11,cd-12} with the additional introduction of \lq\lq fluxes'' arising in the presence of the action of a group $G$. While the technical details of our construction do not apply immediately to infinite chains or quantum field theories, the main important properties are expected to remain valid in the thermodynamic limit, as explained in \cite{cd-11}.

\subsection{Characterisation of the twist operators}
Let us consider a finite dimensional Hilbert space $\mathcal{H}$ describing a bipartition $A \cup \bar{A}$ as in \eqref{Hilbert_space_bipartition}. We take $n$ copies of the above, so that the total Hilbert space is $\mathcal{H}^{\otimes n}$, and we refer to it as the \textit{replica model}.
We aim to define a twist operator $\TT_A$, associated with a cyclic permutation among the replicas restricted to the subsystem $A$. We do so by requiring that on any factorised state of the replica model
\be
\label{def_replica_factorised_state}
\ket{v_1\,,\dots,v_n} \otimes \ket{\bar{v}_1\,, \dots,\bar{v}_n} \equiv \l\otimes_{j=1}^n \ket{v_j}\r\otimes\l\otimes_{j=1}^n \ket{\bar{v}_j}\r\,, \quad \ket{v_j} \in \mathcal{H}_A\,, \ket{\bar{v}_j} \in \mathcal{H}_{\bar{A}},
\ee
the action of $\TT_A$ is\footnote{By linearity this definition allows expressing the action of $\TT_A$ on any vector in $\mathcal{H}^{\otimes n}$.}
\be
\label{T_A definition}
\TT_A \l\ket{v_1\,, v_2\,,\dots,v_n} \otimes \ket{\bar{v}_1\,, \,,\dots,\bar{v}_n}\r \equiv \ket{v_n\,, v_1\,,\dots,v_{n-1}} \otimes \ket{\bar{v}_1\,,\dots,\bar{v}_n}\,.
\ee
Physically, $\TT_A$ implements the permutation $j\rightarrow j+1$ on $A$, and $j$ is a replica index.

In the presence of a global symmetry associated with a group $G$, it is possible to \lq\lq charge'' the twist operator defined above via the action of $G$. Following the terminology of \cite{gs-18}, this corresponds to the additional insertion of Aharonov–Bohm fluxes between the replicas. The action of $G$ on $\mathcal{H}$, defined in \eqref{G_unitary_rep}, is naturally extended to the replica model $\mathcal{H}^{\otimes n}$.

We aim to construct a composite twist operator  $\TT_A^{\{g_1,\dots,g_n\}}$ obtained as the combination of the replica shift $j\rightarrow j+1$ and the insertion of a flux $g_j$ between the $j$th and the $(j+1)$th replicas. We do that, by defining
\be
\TT_A^{\{g_1,\dots,g_n\}} \equiv \TT_A \cdot (\hat{g}_{1,A}\otimes \dots \otimes \hat{g}_{n,A } \otimes \mathbbm{1}^{\otimes n}_{\bar{A}})\,,
\ee
from which it follows that
\be
\TT_A^{\{g_1,\dots,g_n\}}\l\ket{v_1\,,\dots,v_n} \otimes \ket{\bar{v}_1\,,\dots,\bar{v}_n}\r = \hat{g}_{n,A } \ket{v_n}\otimes \hat{g}_{1,A } \ket{v_1}\otimes \dots \otimes \ket{\bar{v}_1\,, \,,\dots,\bar{v}_n}\,,
\ee
where $\hat{g}_{j,A}$ refers to the action of $g_j$ restricted to $A$ and $\mathbbm{1}$ is the action of the identity element $1 \in G$. An operator is therefore associated to any $n$-tuple $\{g_1,\dots,g_n\}$ and, in particular, $\TT_A$ is recovered for $\{g_1,\dots,g_n\} = \{1,\dots, 1\}$. 

In the remaining part of this section, we investigate some useful properties of the composite twist operators. Namely, we relate them to specific traces (charged moments) that appear in the computation of the R\'enyi entanglement asymmetry. Then, we show that distinct twist operators can be related to each other via global unitary transformations induced by the group elements. Finally, we present the mutual locality relations of twist operators with local observables.

\begin{enumerate}
    \item \textbf{Computation of traces.}
    Let $\rho \in \text{End}(\mathcal{H})$ be the density matrix of a (possibly mixed) state and $\rho_A = \Tr_{\mathcal{H}_{\bar{A}}}\rho$. We show that the following identities hold: 
    \be 
    \label{trace identity 1}
    \Tr_{\mathcal{H}^{\otimes n } }\l\rho^{\otimes n} \TT_A\r = \Tr_{\mathcal{H}_A}\rho_A^n\,,
    \ee
    \be
    \label{trace identity 2}
    \Tr_{\mathcal{H}^{\otimes n } }\l\rho^{\otimes n} \TT_A^{\{g_1,\dots,g_n\}}\r = \Tr_{\mathcal{H}_A}\l\rho_A\hat{g}_{n,A }\rho_A\hat{g}_{n-1,A }\dots\rho_A \hat{g}_{1,A}\r\,.
    \ee
    These are relations between the (charged) moments of $\rho_A$ and the expectation values of the (composite) twist operators in the replica model. Let us stress that in the left-hand side the traces are taken over $\mathcal{H}^{\otimes n}$, while in the right-hand side the traces are only over one copy of the subsystem $A$.
    Since the first relation is a special case of the second one (obtained when all the $g_j=1$), we focus on \eqref{trace identity 2}.
    
    A straightforward proof can be provided if one picks two orthonormal bases for $\mathcal{H}_A,\mathcal{H}_{\bar{A}}$ and express the definition of the trace in those bases. With a slight abuse of notation, we denote by $\ket{e_j},\ket{\bar{e}_j}$ the generic basis element of $\mathcal{H}_A,\bar{\mathcal{H}}_A$ in the $j$th replica respectively, and we simply expand
    \begin{align}
    &\Tr_{\mathcal{H}^{\otimes n } }\l\rho^{\otimes n} \TT_A^{\{g_1,\dots,g_n\}}\r \nonumber \\
    = &\sum_{\substack{e_1,\dots, e_n \\ \bar{e}_1, \dots, \bar{e}_n}} \bra{e_1,\dots,e_n}\otimes\bra{\bar{e}_1,\dots,\bar{e}_n} \rho^{\otimes n} \TT_A^{\{g_1,\dots,g_n\}} \ket{e_1,\dots,e_n}\otimes\ket{\bar{e}_1,\dots,\bar{e}_n} \nonumber \\
    = &\sum_{e_1,\dots, e_n}\bra{e_1} \rho_A \hat{g}_{n,A} \ket{e_n}\bra{e_2} \rho_A \hat{g}_{1,A}\ket{e_1}\dots \bra{e_n}\rho_A \hat{g}_{n-1,A} \ket{e_{n-1}} \nonumber \\
    = &\Tr_{\mathcal{H}_A}\l\rho_A\hat{g}_{n,A }\rho_A\hat{g}_{n-1,A }\dots \hat{g}_{2,A}\rho_A \hat{g}_{1,A}\r,
    \end{align}
    which gives precisely \eqref{trace identity 2}.

    \item \textbf{Unitary transformations}

    An important observation is that different composite twist operators $\TT_A^{\{g_1,\dots,g_n\}}$, corresponding to different choices of $\{g_1,\dots,g_n\}$, can be related to each other via global unitary transformations. 
    Specifically, given $h_j \in G$, $j=1,\dots, n$, we show that
    \be
    \label{unitary transformation def}
    \l \hat{h}_1 \otimes \dots \otimes \hat{h}_n \r \TT_A^{\{g_1,\dots,g_n\}} \l \hat{h}_1 \otimes \dots \otimes \hat{h}_n \r^{-1} = \TT_A^{\{g^{\prime}_1,\dots,g^{\prime}_n\}}\,, \quad g_j^{\prime} \equiv h_{j+1}g_j h_j^{-1}\,.
    \ee
    In the relation above, the action of $h_j$ is not restricted to the subsystem $A$. Indeed, the region $\bar{A}$ is not affected by the twist operator appearing in Eq. \eqref{unitary transformation def}, and the combined action of $h_j,h^{-1}_j$ gives the identity on that subsystem.

    To prove \eqref{unitary transformation def} it is sufficient to show that  the left- and right-hand side act in the same way on factorised vectors of $\mathcal{H}^{\otimes n}$. Therefore, for a given state \eqref{def_replica_factorised_state} we just compute
    \begin{align}
     &\l \hat{h}_1 \otimes \dots \otimes \hat{h}_n \r \TT_A^{\{g_1,\dots,g_n\}} \l \hat{h}_1 \otimes \dots \hat{h}_n \r^{-1}\ket{v_1\,,\dots,v_n} \otimes \ket{\bar{v}_1\,, \dots,\bar{v}_n} \nonumber \\  
     =&\l \hat{h}_1 \otimes \dots \otimes \hat{h}_n \r \l\hat{g}_{n,A}\hat{h}^{-1}_{n,A}\ket{v_n} \otimes \dots \otimes \hat{g}_{n-1,A}\hat{h}^{-1}_{n-1,A}\ket{v_{n-1}}\otimes\hat{h}^{-1}_{1,\bar{A}}\ket{\bar{v}_1}\otimes \dots \otimes \hat{h}^{-1}_{n,\bar{A}}\ket{\bar{v}_n}\r \nonumber \\
     =& \l\hat{h}_{1,A}\hat{g}_{n,A}\hat{h}^{-1}_{n,A}\ket{v_n} \otimes \dots \otimes \hat{h}_{n,A}\hat{g}_{n-1,A}\hat{h}^{-1}_{n-1,A}\ket{v_{n-1}}\r\otimes\ket{\bar{v}_1\,, \dots,\bar{v}_n} \nonumber \\
     =& \TT_A^{\{h_2g_1h_1^{-1},\dots,h_1g_nh_n^{-1}\}}\ket{v_1\,,\dots,v_n} \otimes \ket{\bar{v}_1\,, \dots,\bar{v}_n}\,,
    \end{align}
    that is an elementary proof of \eqref{unitary transformation def}.
    
    We discuss a number of remarkable consequences of the equation above. First, the relation $g_j^{\prime} = h_{j+1}g_j h_j^{-1}$ implies that
    \be
    g'_n\,g'_{n-1}\,\dots g'_1 = 1 \quad \text{iff} \quad  g_n\,g_{n-1}\,\dots g_1 = 1\,,
    \ee
    regardless of the choice of $\{h_j\}$. Physically, this means that a twist operator $\TT_A^{\{g_1,\dots,g_n\}}$ with a vanishing net Aharonov-Bohm flux accumulated across the replicas can be unitarily equivalent via \eqref{unitary transformation def}  only to another twist operator $\TT_A^{\{g'_1,\dots,g'_n\}}$ with the same property.
    
    Moreover, such unitary transformation always exists and we explicitly characterise it. To do that, we show that for any $\{g'_i\}$ satisfying $g'_n\dots g'_1 = 1$ there is a $n$-tuple $\{g_j\}$ such that
    \be\label{eq:gg'_rel}
    g_j^{\prime} = g_{j+1} g_j^{-1}\,.
    \ee
    A solution to \eqref{eq:gg'_rel} is indeed
    \be
     g_1 = 1\,, \quad g_2 = g_1^\prime\,, \quad g_3= g_2^\prime g_1^\prime\,, \quad \dots \quad g_n= g_{n-1}^\prime\dots g_1^\prime\,.
    \ee
    Therefore, with the previous choices of $\{g_j\}\,,\{g'_j\}$, it holds
    \be
    \label{unitary transformation simplified}
    \l \hat{g}_1 \otimes \dots \otimes \hat{g}_n \r \TT_A \l \hat{g}_1 \otimes \dots \otimes\hat{g}_n \r^{-1} = \TT_A^{\{g^{\prime}_1,\dots,g^{\prime}_n\}}\,,
    \ee
    which means that $\TT_A$ is always unitarily equivalent to a given composite twist operator with a vanishing net flux.
    
    Finally, we notice that distinct unitary transformations can give rise to the same twist operator, since the Eq. \eqref{eq:gg'_rel} has multiple solutions. For example, it is easy to show that
    \be
    \label{global neutrality general}
        \l \hat{g}^{\otimes n}\r \TT_A \l \hat{g}^{\otimes n}\r^{-1} = \TT_A\,,
    \ee
    for any choice of $g \in G$, which means that the standard twist operator $\TT_A$ is neutral under a global unitary transformation. More generally, the Eq. \eqref{eq:gg'_rel} has precisely $|G|$ solutions: indeed, if $\{g_j\}$ is a solution of \eqref{eq:gg'_rel} for a given $n$-tuple $\{g'_j\}$, then also $\{g_jg\}$ satisfies the same relation for any $g \in G$.

    \item \textbf{Action on local observables}

    The last properties we show are the commutation relations between the twist operators and the local observables of the system. This is particularly useful to make an explicit connection between the construction proposed above and the defining properties of twist operators in QFT based on the algebra of local observables \cite{cdmsc-22}.

    Let $\mathcal{O} \in \text{End}(\mathcal{H}_A)$ be an observable of the subsystem $A$. $\mathcal{O}$ can be naturally embedded in the space of observables of the whole system $A\cup \bar{A}$ by mapping it to $\mathcal{O} \otimes \mathbbm{1}_{\bar{A}} \in \text{End}(\mathcal{H})$. Similarly, we can embed $\mathcal{O}$ in the $j$th replica of the replica model defining
    \be
    \mathcal{O}^j \equiv \underbrace{\mathbbm{1}_A\otimes \dots \mathcal{O}}_{j}\otimes\dots \mathbbm{1}_A \otimes \mathbbm{1}^{\otimes n}_{\bar{A}} \in \text{End}(\mathcal{H}^{\otimes n}).
    \ee
    As a straightforward consequence of the definitions, it is not difficult to show that
    \be
    \TT_A \mathcal{O}^j = \mathcal{O}^{j+1}\TT_A,
    \ee
    and we say that the twist operator act as a replica shift $j\rightarrow j+1$ on the observables $\mathcal{O}$ of $A$. Similarly, one has
    \be
    \TT_A^{\{g_1,\dots,g_n\}} \mathcal{O}^j =  \l\hat{g}_j^{\otimes n}\r\mathcal{O}^{j+1}\l\hat{g}_j^{\otimes n}\r^{-1}\TT_A^{\{g_1,\dots,g_n\}}.
    \ee
    Physically, this means that the action of the composite twist operators on a local observable of $A$ amounts to the usual replica shift followed by the action of the group element $g_j$ on $\mathcal{O}$.

Vice versa, local observables of $\bar{A}$ are not affected by the presence of the twist operators. For instance, given $\mathcal{O} \in \text{End}\l \mathcal{H}_{\bar{A}}\r$, and $\mathcal{O}^j$ its embedding in the replica model, one easily shows
\be
    \TT_A^{\{g_1,\dots,g_n\}} \mathcal{O}^j =  \mathcal{O}^j \TT_A^{\{g_1,\dots,g_n\}}.
\ee
Therefore, as expected, the action of the twist operator on $A$ commutes with the observables of the complement $\bar{A}$.

\end{enumerate}

\subsection{Entanglement asymmetry via twist operators}

In this section, we make a connection between the twist operators introduced previously and the definition of the R\'enyi entanglement asymmetry in Eq. \eqref{eq:Ent_asymm}.

To do so, we characterise the R\'enyi asymmetry of a state $\rho \in \text{End}\l \mathcal{H}_A \otimes \mathcal{H}_{\bar{A}}\r$ for an integer $n\geq 2$. First, we recall that \eqref{trace identity 1} allows expressing the R\'enyi entropy of $\rho_A$ via the twist operator. Then, by making use of the definition of $\tilde{\rho}_A$ in Eq. \eqref{eq:rho_tilde} and after a change of variable, we get
\be\label{eq:rho_tild_ch_var}
\begin{split}
\Tr\l\tilde{\rho}_A^n\r &= \frac{1}{|G|^n}\sum_{g_1,\dots,g_n \in G}\Tr\l \hat{g}_{1,A}\rho_A \hat{g}_{1,A}^{-1}\hat{g}_{2,A}\rho_A \hat{g}_{2,A}^{-1}\dots \hat{g}_{n,A}\rho_A \hat{g}_{n,A}^{-1}\r \\
&=   \frac{1}{|G|^{n-1}}\sum_{g'_1,\dots,g'_{n-1} \in G}\Tr\l \rho_A \hat{g}'_{1,A}\dots \rho_A \hat{g}'_{n-1,A}\rho_A (\hat{g}'_{1,A},\dots \hat{g}'_{n-1,A})^{-1}\r.
\end{split}
\ee
Here, the second line is obtained by introducing a change of variable $g'_j = g^{-1}_{j} g_{j+1}$. Notably, the $g'_j$ are not independent since they satisfy the constraint $\prod_j g'_j = 1$. This allows us to express $\Tr(\tilde{\rho}_A^n)$ as a sum of $|G|^{n-1}$ terms, as shown in Eq. \eqref{eq:rho_tild_ch_var}.
Finally, we employ the property \eqref{trace identity 2} and we express the moments of $\tilde{\rho}_A$ via the composite twist operators as
\be
\label{Tr_rho_tilde_simplified}
 \Tr\l\tilde{\rho}_A^n\r = \frac{1}{|G|^{n-1}}\sum_{g_1,\dots,g_{n-1} \in G}\Tr\l\rho^{\otimes n} \TT_A^{\{g_1,\dots,g_{n-1},(g_1,\dots,g_{n-1})^{-1}\}}\r,
\ee
and, eventually, from \eqref{eq:Ent_asymm} we get
\be\label{eq:asym_t_op}
\Delta S_n = \log |G| + \frac{1}{1-n}\log\l  \sum_{g_1,\dots,g_{n-1} \in G} \frac{\Tr\l\rho^{\otimes n} \TT_A^{\{g_1,\dots,g_{n-1},(g_1,\dots,g_{n-1})^{-1}\}}\r}{\text{Tr}\l \rho^{\otimes n} \TT_A\r}\r.
\ee
We point out that all the possible $|G|^{n-1}$ composite twist operators with a vanishing net Aharonov–Bohm flux appear explicitly in the sum in \eqref{eq:asym_t_op}. 
Moreover, it is worth noting that \eqref{eq:asym_t_op} is invariant after a rescaling
\be
\TT_A\rightarrow \lambda \TT_A \,, \quad \TT_A^{\{g_1,\dots,g_{n-1},(g_1,\dots,g_{n-1})^{-1}\}} \rightarrow \lambda \TT_A^{\{g_1,\dots,g_{n-1},(g_1,\dots,g_{n-1})^{-1}\}}\,.
\ee
This last observation is fundamental for quantum field theories (QFTs) in which spontaneous symmetry breaking occurs, as we explain below.

First, one can argue that, due to the global unitary transformation in Eq. \eqref{unitary transformation simplified}, the charged and uncharged twist operators must have the same (UV) scaling dimension. For instance, in the limit of a small region, their correlation functions are expected to be the same as those computed at the UV critical point. Since the critical point is invariant under the symmetry group $G$, these expectation values are equal.

Also, because of Eq. \eqref{unitary transformation simplified}, the normalization of $\mathcal{T}_A$ unambiguously determines the normalization of $\TT_A^{{g_1,\dots,g{n-1},(g_1,\dots,g_{n-1})^{-1}}}$. Consequently, the ratio appearing in \eqref{eq:asym_t_op}, and therefore the entanglement asymmetry, is expected to be universal. It is important to emphasize that this is not the case for entanglement entropy (see Refs. \cite{ot-15,cw-94}) due to the presence of UV divergences.

\section{Form factor program for the twist fields in the Ising ferromagnetic phase}\label{sec:FF_Ising}

In this section, our focus is on the integrable QFT of the Ising model in 1+1 dimensions in its ferromagnetic phase. We aim to provide an explicit characterisation of the twist operators introduced in Sec. \ref{sec:twist_op} by means of form factor bootstrap. In particular, we consider $n$ replicas of the Ising model, and we investigate a family of branch-point twist fields. This allows us to obtain the R\'enyi entropy of the symmetry-broken phase, which differs explicitly from the one of the disordered phase investigated in \cite{ccd-08}.

Before entering into the details of the Ising model, it is worth making a connection between the twist operators of Sec. \ref{sec:twist_op} and the twist fields considered in the literature, widely used in 1+1 dimensional field theories (see e.g. Ref. \cite{cc-09} for CFTs, and Ref. \cite{ccd-08} for massive integrable QFTs). For any region $A =(x,\infty)$, representing a half-line originating from $x$ and extending infinitely, we regard the corresponding twist operator as a (semi-local) field inserted at $x$ and we denote it by $\mathcal{T}(x)$, namely
\be
\mathcal{T}(x) \sim \mathcal{T}_A\,, \quad A =(x,\infty)\,,
\ee
up to a non-universal proportionality constant that has been neglected. These fields are the building blocks needed to reconstruct the entanglement properties of any region, and through their insertion at different points, one can express the twist operator of any union of disjoint intervals \cite{cc-09}. A paradigmatic example is the case of a single interval, where it is possible to express
\be
\mathcal{T}_A \sim \mathcal{T}(0)\mathcal{T}^\dagger (\ell)\,, \quad A = (0,\ell)\,,
\ee
with $\mathcal{T}\,,\mathcal{T}^\dagger$ a pair of Hermitian conjugated twist fields. This correspondence has also been generalised to composite twist fields in Refs. \cite{cdmsc-22, cmc-23}.

Despite the extensive literature existing on twist fields in 1+1 dimensional integrable QFT, the vast majority of those works \cite{cc-09,Doyon-09,cl-11,CAlvaredo-17,ch-21} analyse the paramagnetic (disordered) phase only, where a single vacuum is present, and its entanglement is characterised. However, in the presence of spontaneous symmetry breaking, multiple vacua are present, and one must be cautious when specifying them. Moreover, as there might be elementary excitations (kinks) interpolating between different vacua, analytic continuations in the rapidity space may result in a mixing of the corresponding form factors, a mechanism that is well understood for the $q$-state Potts field theory \cite{dc-98}, which describes the Ising model for $q=2$. To the best of our knowledge, the consequence of vacuum degeneracy in the context of entanglement for field theories has not been explored yet. Our objective is to address this gap by examining the simplest, yet non-trivial case: the Ising model. In particular, in this section we aim to formulate and solve the form factor bootstrap equations for the standard twist fields in the ferromagnetic case. These results will be eventually employed to compute the R\'enyi entropy of a symmetry-broken ground state.

We first review some basic properties of the Ising theory \cite{Delfino-98}, and then we discuss its $n$-replica model, where the twist fields arise. The theory can be regarded as a relevant perturbation of the Ising CFT \cite{dms-97, z-89} with Euclidean action
\be
 \text{S} = \text{S}_{\text{CFT}} + \lambda \int dx d\tau \ \varepsilon(x,\tau)\,.
\ee
Here $\text{S}_{\text{CFT}}$ is the action at criticality, associated to a CFT with central charge $c=1/2$, $\lambda$ is a parameter with the dimension of a mass and $\varepsilon(x,\tau)$ is a scalar field of dimension $1$ representing an energy density. Depending on the sign of $\lambda$, the theory is in its paramagnetic (disordered, $\lambda<0$) or ferromagnetic (ordered, $\lambda>0$) phase. We focus on the ferromagnetic phase where spontaneous symmetry breaking arises, and two degenerate vacua $\ket{+},\ket{-}$ related by $\mathbb{Z}_2$ symmetry are present. The elementary excitations above the ground state(s) are multi-kink configurations interpolating between the two vacua \cite{dc-98}. In particular, a single-particle (kink) state with rapidity $\theta$, denoted by
\be
\ket{K_{+-}(\theta)}\,,
\ee
interpolates between $\ket{-}$ at $-\infty$ and $\ket{+}$ at $\infty$. In general, a configuration with $N$ kinks has the form
\be
\ket{K_{\alpha_1 \alpha_2}(\theta_1)K_{\alpha_2 \alpha_3}\dots (\theta_2)\dots K_{\alpha_{N} \alpha_{N+1}}(\theta_N)}\,,
\label{eq:multi_kink}
\ee
with $\alpha_j \in \{+,-\}$ labeling the vacua and $\{\theta_j\}_{j=1,\dots,N}$ is a set of rapidities. We say that the configuration is neutral, or topologically trivial, if $\alpha_1=\alpha_{N+1}$. In general, it is convenient to identify four topological sectors, labeled by the values of $(\alpha_1,\alpha_{N+1})$: physically, these sectors correspond to different choices of the boundary conditions (i.e. spin up or spin down) at $x=-\infty$ and $x= +\infty$, and they can be regarded as constraints for the states interpolating between them. 

The multi-kink configurations are the (asymptotic) excited states of the theory, and their energy-momentum is
\be
(E,P) = \l m\sum^{N}_{j=1}\cosh \theta_j\,,m\sum^{N}_{j=1}\sinh \theta_j\r\,.
\ee
Here $m = |\lambda|$ is the mass of each kink, and it plays the role of inverse correlation length away from criticality.
When two kinks scatter, they get a scattering phase $-1$, which does not depend on the rapidity difference, a property that is ultimately related to the equivalence (see \cite{dms-97}) between the Ising model and free fermionic Majorana field theory\footnote{There are some technical differences between the Ising model and the field theory of free Majorana fermions, which can be traced back to the presence of distinct spin sectors (see Ref. \cite{dms-97} for details). However, since this feature does not play any role in this work, we do not discuss it further.}.

\begin{figure}[t]
    \centering
	\includegraphics[width=0.67\linewidth]{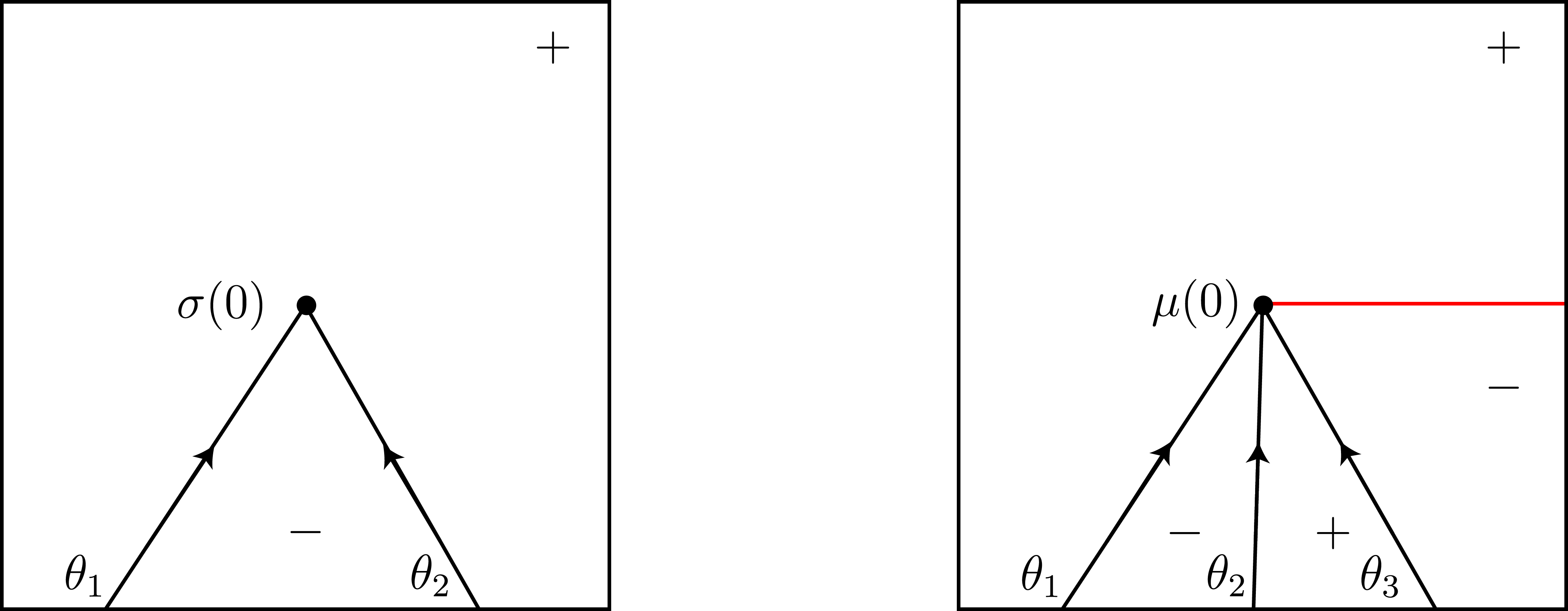}
    \caption{(Left) Order field $\sigma(x)$: the two-particle form factor $\bra{+}\sigma(0)\ket{K_{+-}(\theta_1)K_{-+}(\theta_2)}$ is shown. (Right) Disorder field $\mu(x)$: the three-particle form factor $\bra{+}\mu(0)\ket{K_{+-}(\theta_1)K_{-+}(\theta_2)K_{+-}(\theta_3)}$ is represented. The red line corresponds to a spin-flip exchanging the two vacua.}
    \label{fig:ord_dis}
\end{figure}

For any field $\mathcal{O}(x)$ one can define the form factors, i.e. the matrix elements of $\mathcal{O}$ between the vacua and the multi-kink states, as
\be
\bra{\pm} \mathcal{O}(0)\ket{K_{\alpha_1 \alpha_2}(\theta_1)K_{\alpha_2 \alpha_3} (\theta_2)\dots K_{\alpha_{N} \alpha_{N+1}}(\theta_N)}\,.
\ee
In principle, depending on the field $\mathcal{O}$, many non-vanishing form factors might be present; however, some of them vanish identically due to topological constraints, as discussed in Ref. \cite{dc-98}. Specifically, for the order field $\sigma$ only the form factors with neutral states may be non-vanishing, and they are
\be
\bra{\pm} \sigma(0)\ket{K_{\pm \alpha_2}(\theta_1)K_{\alpha_2 \alpha_3}(\theta_2),\dots K_{\alpha_{N} \pm}(\theta_N)}\,,
\ee
together with the vacuum expectation values (VEVs) $\bra{\pm}\sigma(0)\ket{\pm}$. In contrast, the disorder field $\mu(0)$ is topologically non-trivial, as it introduces a spin-flip along the half-line $x \in [0,+\infty)$ that mixes the two vacua, and the form factors which are not ruled out under topological constraints are
\be
\bra{\pm} \mu(0)\ket{K_{\pm \alpha_2}(\theta_1)K_{\alpha_2 \alpha_3} (\theta_2)\dots K_{\alpha_{N} \mp}(\theta_N)}\,.
\ee
In Fig. \ref{fig:ord_dis} we represent pictorially the form factors of the order and disorder field. While $\sigma(x)$ is a local field, $\mu(x)$ is semi-local, as it acts as a spin-flip over $(x,\infty)$ (depicted as a red line in Fig. \ref{fig:ord_dis}). The  latter properties fix the boundary conditions at infinity for incoming states interpolating with the outgoing vacuum. In particular, only an even/odd number of kinks gives rise to non-vanishing form factors for $\sigma(x),\mu(x)$ respectively. 

In the next sections, we first extend the analysis above to the twist fields of the replica model, identifying their topological constraints. Then, via the bootstrap program, we characterise analytically their non-vanishing form factors, and we give an analytical expression for the two-particle ones.

\subsection{Standard twist fields}

We take $n$ decoupled copies of the Ising model, and we consider the resulting model itself as a QFT, which we refer to as the $n$-replica model. The (asymptotic) spectrum of the replica theory is expressed in terms of single-replica vacua and kink states, as no correlations are present between distinct replicas. In particular, in the ordered phase, there are $2^n$ degenerate vacua which are obtained as tensor products of single-replica vacua:
\be
\ket{+}^{\otimes n},\, \ket{+}^{\otimes n-1}\otimes\ket{-},\, \dots,\,\ket{-}^{\otimes n}\,.
\ee
The excitations can be described via the insertion of kinks in each replica, similarly to what happens with quasi-particle excitations in the paramagnetic phase (see \cite{ccd-08}). Moreover, the scattering matrix between two kinks at replicas $i$ and $j$ with rapidities $\theta_i$, $\theta_j$ respectively is
\be
\label{scattering kinks}
S_{i j}(\theta_i-\theta_j) = \begin{cases} -1 \quad i=j \,,\\ 
1 \quad i \neq j\,,
\end{cases}
\ee
since no interactions between distinct replicas are present. This further implies that two kinks on the same replica satisfy the Faddeev-Zamolodchikov algebra:
 \be
 \label{FZ_algebra}
    K_{\pm \mp}(\theta_1)K_{\mp \pm}(\theta_2) = -K_{\pm \mp}(\theta_2)K_{\mp \pm}(\theta_1),
\ee
which is consistent with the algebra of the scattering kinks in the $q$-state Potts model \cite{chim1992integrable,dc-98} for $q=2$.

The standard branch-point twist field can be introduced as a semi-local field of the replica model implementing the replica cyclic permutation $j\rightarrow j+1$. In particular, we denote this field by $\mathcal{T}(x)$ and require that
\be
\mathcal{T}(x)\mathcal{O}_j(y) = \begin{cases} \mathcal{O}_{j+1}(y)\mathcal{T}(x) \quad x<y\,,\\ 
\mathcal{O}_{j}(y)\mathcal{T}(x) \quad \text{otherwise}\,,
\end{cases}
\label{eq:comm_st_Tfields}
\ee
for any local field\footnote{
One should be careful about the definition of local fields, and more in general local observables. Here, we just mean that $\mathcal{O}$ belongs to the algebra generated by the fields $\varepsilon$ (the energy density) and the order operator $\sigma$. This requirement is very natural in the lattice counterpart, as it corresponds to the usual notion of locality in the computational basis of the quantum spin chain.
} $\mathcal{O}_j(x)$ inserted in the $j$th replica. According to this commutation relation (see Refs. \cite{cc-09, ccd-08}), we say that $\mathcal{T}(x)$ inserts a branch cut on the semi-infinite line $(x,\infty)$, which connects the $j$th replica with the $(j+1)$th. It is important to stress that Eq. \eqref{eq:comm_st_Tfields}, rather than unambiguously identifying a single field $\mathcal{T}(x)$, selects a space of fields with given monodromy properties (details are found in Ref. \cite{bd-17}). However, we are interested in the lightest fields satisfying \eqref{eq:comm_st_Tfields}, which correspond to a deformation of the primary twist fields of CFTs \cite{cc-09}. These are scalar fields with scaling dimension
\be
\Delta = \frac{c}{12}\l n -\frac{1}{n} \r,
\ee
with $c$ the central charge of the theory ($c=1/2$ for the Ising CFT). These fields are being studied extensively in the paramagnetic phase of the Ising field theory, and we refer the reader to Refs. \cite{ccd-08,cl-11}.

Here, we study the form factors of branch-point twist fields in the ferromagnetic phase of the Ising model. These are matrix elements of $\TT$, $\TT^{\dagger}$ between the multi-kink state and one of the $2^n$ vacua. For later purposes, the most interesting form factors are the ones with the vacuum $\bra{+}^{\otimes n}$. However, as we will see, other form factors are generated from the latter using the bootstrap equations. 

\subsubsection{Zero-particle form factors}
\begin{figure}[t]
    \centering
	\includegraphics[width=0.76\linewidth]{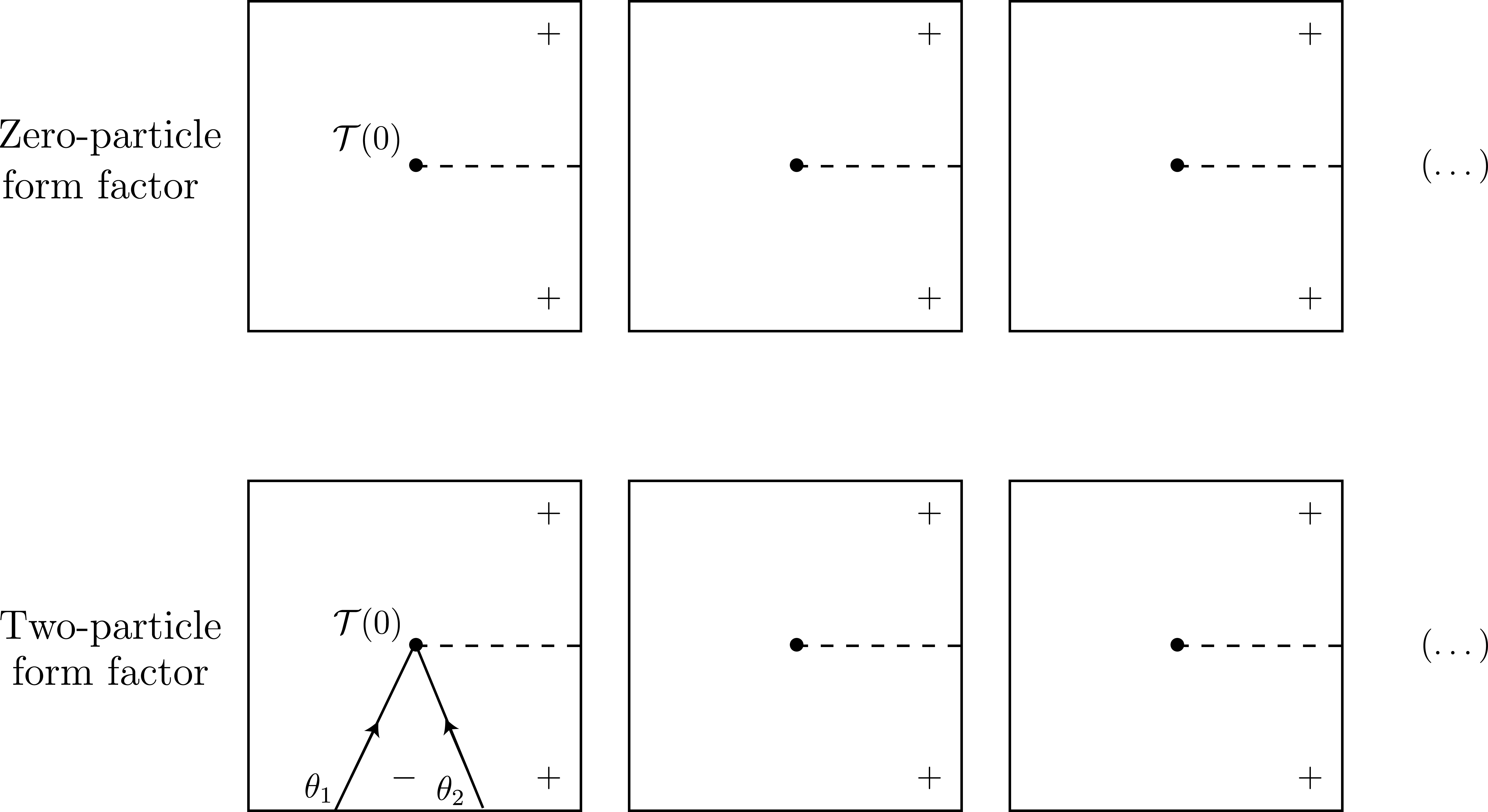}
    \caption{The form factors of the twist field $\mathcal{T}(0)$. Consecutive replicas are connected to each other via the branch cut, represented by a dashed line starting at $x=0$.
    Top: vacuum expectation value ${}^{\otimes n}\bra{+}\mathcal{T}(0)\ket{+}^{\otimes n}$.
    Bottom: Two-particle form factors ${}^{\otimes n}\bra{+}\mathcal{T}(0)\ket{K_{+-}(\theta_1)K_{-+}(\theta_2),+,\dots,+}$ with two kinks in the same replica.
    }
    \label{fig:ff_tfield}
\end{figure}

The simplest form factor that is expected to be non-vanishing is the VEV
\be
{}^{\otimes n}\bra{+}\mathcal{T}(0)\ket{+}^{\otimes n}\,.
\ee
We represent it pictorially in Fig. \ref{fig:ff_tfield}: the $n$ replicas are connected together along the branch cuts inserted by the twist fields $\mathcal{T}$ at the origin of the space-time, the state $\ket{+}^{\otimes n}$ (${}^{\otimes n}\bra{+}$) is the incoming (outgoing) state represented below (above) the origin in each replica.
In particular, since the standard twist fields exchange replicas but do not act as spin-flips, if $\pm$ is present in the $j$th replica above the branch cut, then $\pm$ has to be present also in the $(j+1)$th replica just below the branch cut. 
In other words, fixing the boundary condition of the outgoing state at spacial infinity unambiguously fixes the boundary conditions of the incoming state. Specifically, the boundary conditions at $x=-\infty$ remain the same for both outgoing and incoming states, while the boundary conditions at $x=+\infty$ are connected through a replica shift $j\rightarrow j+1$. For a similar discussion concerning the (single-replica) $q$-state Potts model, we refer the reader to Ref. \cite{dc-98}.

According to these considerations, it is not difficult to show that the only other non-vanishing zero-particle form factor is
\be
{}^{\otimes n}\bra{-}\mathcal{T}(0)\ket{-}^{\otimes n}\,.
\ee
Its value can be related to the one of ${}^{\otimes n}\bra{+}\mathcal{T}(0)\ket{+}^{\otimes n}$ via a global $\mathbb{Z}_2$ symmetry. Indeed, let us consider the global unitary transformation $\mu$ \footnote{Notice that the operator $\mu$ acts as a spin-flip everywhere while the field $\mu(x)$ acts on the half-line $(x,+\infty)$. Roughly, the relation between these operators is (up to a proportionality constant) $\mu \propto \mu(-\infty)\mu^{\dagger}(+\infty)$.
} which generates the $\mathbb{Z}_2$ symmetry exchanging the two vacua. In the replica theory, it holds
\be
\mu^{\otimes n} \ket{\pm}^{\otimes n} = \ket{\mp}^{\otimes n}.
\ee
Since the twist field is neutral with respect to the action of the global spin-flip $\mu^{\otimes n}$ (see Eq. \eqref{global neutrality general}), this immediately implies that the two zero-particle form factors above are equal: this value is referred to as
\be\label{TT vev}
\tau \equiv {}^{\otimes n}\bra{\pm}\mathcal{T}(0)\ket{\pm}^{\otimes n}.
\ee

\subsubsection{Vanishing one-particle form factors and further constraints}\label{sec:one_p_ff}

Here, we show that the one-particle form factors are all vanishing. To do that, it is sufficient to observe that these form factors are not compatible with the topological constraints arising from the boundary conditions at spacial infinity.

Let us consider a generic outgoing vacuum state $\bra{\alpha_1,\dots,\alpha_n}$, with $\alpha_j = \pm$. If $\mathcal{T}(0)$ interpolates between a multi-particle incoming state and the vacuum above, we can immediately obtain the boundary conditions for the incoming state. Indeed, it is easy to show that the boundary conditions for the latter in the $j$th replica at $x=-\infty$, $x=\infty$ have to be $\alpha_{j}$, $\alpha_{j-1}$ respectively. This property rules out the presence of matrix elements with an odd number of kinks having $\bra{\alpha_1,\dots,\alpha_n}$ as outgoing state, and, in particular, the one-particle form factor.

Further constraints are also implied, and, for the sake of concreteness, we only discuss them for the outgoing vacuum $\bra{+,\dots,+}$. Since $+$ should be present at $x=\pm \infty$ of the incoming state in the $j$th replica, only an even number of kinks has to be present for any replica. For example, two-particle states with two kinks inserted at different replicas have vanishing form factors. The latter property is specific to the ferromagnetic phase of the Ising model, while the same mechanism does not arise in the paramagnetic one \cite{ccd-08}.

\subsection{Two-particle form factors}\label{sec:2p_ff}

The two-particle form factors of $\mathcal{T}(x)$ are responsible for the leading corrections to the saturation value of the R\'enyi entropy \cite{ccd-08}. They are generically non-vanishing, as they are not ruled out by symmetry or topological constraints, and can be exactly determined via the bootstrap equations. We perform an analysis, similar to Ref. \cite{ccd-08} where the paramagnetic phase was considered: analogies arise, due to the same scattering properties, but important conceptual differences are present, due to the vacuum degeneracy. In analogy with \cite{ccd-08}, the two-kink form factors can be determined by their monodromy conditions and the residues at their kinematic poles. We shall first derive the monodromy equations, and then move to the equations for the kinematic poles.

Let us focus on the following two-kink form factor
   \be\label{eq:F_11}
    F_{11}^{(n)}(\theta_{1}-\theta_2) \equiv \prescript{\otimes n}{}{\bra{+}}\mathcal{T}(0)\ket{K_{+-}(\theta_1)K_{-+}(\theta_2),+,\dots,+},
    \ee
which depends on the difference $\theta_1-\theta_2$ only, due to relativistic invariance. In general, this is expected to be non-vanishing, since, as argued in Sec. \ref{sec:one_p_ff}, two kinks are presented in the same replica (the first). We give a pictorial representation of $F^{(n)}_{11}(\theta_1-\theta_2)$ in Fig. \ref{fig:ff_tfield}.

As a consequence of replica symmetry there is no explicit dependence on the replica $j$ where the pair of kinks is inserted, and we have
\be
    F_{jj}^{(n)}(\theta_{1}-\theta_2) \equiv \prescript{\otimes n}{}{\bra{+}}\TT(0)\ket{\underbrace{+,\dots,+,K_{+-}(\theta_1)K_{-+}(\theta_2)}_j,+,\dots,+} = F_{11}^{(n)}(\theta_{1}-\theta_2)\,,
    \ee
with $j=1,\dots,n$.

Let us discuss the \textit{monodromy equations} of $F^{(n)}_{11}(\theta)$. The rapidity shift $\theta_1\rightarrow \theta_1+2\pi i$ in Eq. \eqref{eq:F_11} has the effect of moving the corresponding kink from the first to the second replica. In addition, the topological sectors are mixed with each other, and we get
\begin{align}
    \label{monodromy general j shift}
    &\prescript{\otimes n}{}{\bra{+}}\mathcal{T}(0)\ket{K_{+-}(\theta_1 + 2\pi i)K_{-+}(\theta_2),+,\dots,+} \nonumber \\ = &\bra{-,+,\dots,+}\TT(0)\ket{K_{-+}(\theta_2),K_{+-}(\theta_1),+,\dots,+}\,.
\end{align}
This means that the outgoing vacuum is modified under the rapidity shift above. This is not particularly surprising, since we already argued in Sec. \ref{sec:one_p_ff} that the two-particle form factors of $\bra{+,\dots,+}$ with kinks in distinct replicas vanish. 

Similarly, for $j=1,\dots,n-1$, we have
\begin{align}
    &\prescript{\otimes n}{}{\bra{+}}\mathcal{T}(0)\ket{K_{+-}(\theta_1 + 2\pi ij)K_{-+}(\theta_2),+,\dots,+} \nonumber \\ = &\bra{\underbrace{-,\dots,-}_{j},+,\dots,+}\TT(0)\ket{\underbrace{K_{-+}(\theta_2),-,\dots,-, K_{+-}(\theta_1)}_{j+1},+,\dots,+}\,,
    \end{align}
and the kink moves from the first to the $(j+1)$th replica under $\theta_1\rightarrow \theta_1+2\pi i j$. If the one performs the shift $\theta_1 \rightarrow \theta_1+2\pi i n$, the kink goes back to the first replica, but all the signs are exchanged, namely
\begin{align}
    &\prescript{\otimes n}{}{\bra{+}}\mathcal{T}(0)\ket{K_{+-}(\theta_1 + 2\pi in)K_{-+}(\theta_2),+,\dots,+} \nonumber \\ = &\prescript{\otimes n}{}{\bra{-}}\mathcal{T}(0)\ket{K_{-+}(\theta_2)K_{+-}(\theta_1),-,\dots,-}\,. 
    \end{align}
Using further the $\mathbb{Z}_2$ neutrality of the twist fields, and the (fermionic) scattering properties \eqref{FZ_algebra}, we eventually end up with
\begin{align}
    &\prescript{\otimes n}{}{\bra{+}}\mathcal{T}(0)\ket{K_{+-}(\theta_1 + 2\pi in)K_{-+}(\theta_2),+,\dots,+} \nonumber \\ = & - \prescript{\otimes n}{}{\bra{+}}\mathcal{T}(0)\ket{K_{+-}(\theta_1)K_{-+}(\theta_2),+,\dots,+}\,, 
\end{align}
that is $F^{(n)}_{11}(\theta+2in\pi) = -F^{(n)}_{11}(\theta)$. The monodromy equations thus lead to an anti-periodicity condition on the fundamental two-kink form factor.

Following closely the arguments in \cite{Doyon-09}, the form factor $F_{11}^{(n)}(\theta)$ has two \textit{kinematic poles} at $\theta=i\pi$ and $\theta= 2\pi i n - i\pi$. Their residues are
    \be
    \text{Res}_{\theta= i\pi}F_{11}^{(n)}(\theta) = i \tau     \,, \quad \text{Res}_{\theta= (2n-1)i\pi}F_{11}^{(n)}(\theta) = -i\tau\,,
    \ee
    with $\tau$ the vacuum expectation value \eqref{TT vev}. Combining all these results, we have that both the monodromy and the bootstrap equations of $F_{11}^{(n)}(\theta)$ are the same encountered in the paramagnetic phase \cite{ccd-08}. Therefore, a solution\footnote{It is well-known that the bootstrap equations have multiple solutions, but the $\Delta$-theorem \cite{dsc-96} provides a severe additional constraint. Here, as in \cite{ccd-08}, the solution with the mildest growth for $|\theta|\rightarrow \infty$ is given. This one is expected to describe the twist field with the lowest scaling dimension, corresponding to a primary field of the associated CFT.
    } to the bootstrap can be easily provided:
    \be
    \label{two-kink form factor final}
    F_{11}^{(n)}(\theta) = \frac{i\tau \,\cos\l\frac{\pi}{2n}\r\sinh\l\frac{\theta}{2n}\r}{n\,\sinh\l\frac{\theta-i\pi}{2n}\r\sinh\l\frac{\theta+i\pi}{2n}\r}\,.
    \ee

\subsection{R\'enyi entropy and its analytic continuation} \label{sec:entropy_analytic_continuaiton}

With the previous results, we can finally provide a form-factor expansion for the two-point function of the twist fields, valid in the limit $m\ell \gg 1$. This expansion is then used to obtain the R\'enyi entropies. Eventually, we analytically continue the replica index $n\rightarrow 1$ to get the entanglement entropy.

Starting from the replica state $\ket{+}^{\otimes n}$, the quantity we need to compute is
\be
\text{Tr}\l \rho^{\otimes n} \TT_A\r = \prescript{\otimes n}{}{\bra{+}} \TT(0)\TT^\dagger(\ell)\ket{+}^{\otimes n}\,.
\ee
We insert a resolution of the identity between the branch-point twist fields and, at the two-particle approximation, we get
\be\label{eq:tfield_2point}
\prescript{\otimes n}{}{\bra{+}} \TT(0)\TT^\dagger(\ell)\ket{+}^{\otimes n} \simeq |\tau|^2 + \sum_{j=1}^n \int_{\theta_1 > \theta_2} \frac{\mathrm{d}\theta_1\,\mathrm{d}\theta_2}{(2\pi)^2} |F_{jj}^{(n)}(\theta_1-\theta_2)|^2 e^{-m\ell\l\cosh\theta_1 + \cosh\theta_2\r}\,.
\ee
In deriving the expression above, we used the fact that the only vacuum that can be connected to ${}^{\otimes n}\bra{+}$ by $\mathcal{T}(0)$
is $\ket{+}^{\otimes n}$. Moreover, only pairs of kinks inserted in the same replica $j$ give a non-vanishing contribution in the expansion above.

Performing a change of variable $\theta = \theta_1-\theta_2$, $\Theta = \frac{\theta_1+\theta_2}{2}$ and using the integral representation of the Bessel function \cite{gr-14}
\be
K_0(z) = \int^{\infty}_0 \mathrm{d}x\, e^{-z\cosh x}\,,
\ee
we eventually get
\be\label{eq:two_p_tfield}
\prescript{\otimes n}{}{\bra{+}} \TT(0)\TT^\dagger(\ell)\ket{+}^{\otimes n} \simeq |\tau|^2 + \frac{n}{4\pi^2}\, \int^{+\infty}_{-\infty} \mathrm{d}\theta\,K_0\l 2 m \ell \cosh\l\frac{\theta}{2}\r\r |F_{11}^{(n)}(\theta)|^2,
\ee
as $|F_{jj}^{(n)}(\theta)|^2 = |F_{11}^{(n)}(\theta)|^2$ does not depend on $j$ explicitly.
In the large volume limit $m \ell \rightarrow \infty$ the correlation function factorises and the square of the vacuum expectation value $|\tau|^2$ is recovered in Eq. \eqref{eq:tfield_2point}: the first non-trivial correction to it, which decays exponentially to zero as $\sim e^{-2m\ell}$, comes from the second term of Eq. \eqref{eq:two_p_tfield} and is due to the pairs of kinks. This result should be compared to the (two-particle approximation of) the two-point function obtained in the paramagnetic phase \cite{ccd-08}. The crucial difference is that in the paramagnetic phase form factors with single-particle excitations inserted in distinct replicas are allowed, which amounts to replace (see Ref. \cite{ccd-08})
\be
|F_{11}^{(n)}(\theta)|^2 \rightarrow \sum^{n-1}_{j=0}|F_{11}^{(n)}(-\theta+2\pi i j)|^2
\ee
in Eq. \eqref{eq:two_p_tfield}. As we show below, this difference has important consequences in the replica limit $n\rightarrow 1$.

Now, we express the R\'enyi entropy of $\ket{+}$ as \cite{ccd-08}
\begin{align}\label{eq:Sn_Ising}
S_n &= \frac{1}{1-n}\log \prescript{\otimes n}{}{\bra{+}} \TT(0)\TT^\dagger(\ell)\ket{+}^{\otimes n} + \dots \nonumber \\
&\simeq \frac{1}{1-n}\log\l 1+\frac{n}{4\pi^2 }\, \int^{+\infty}_{-\infty} \mathrm{d}\theta\,K_0\l 2 m \ell \cosh\l\frac{\theta}{2}\r\r \frac{|F_{11}^{(n)}(\theta)|^2}{|\tau|^2}\r+\dots,
\end{align}
where an irrelevant additive constant, eventually removed by a proper choice of the twist field normalisation, has been neglected. In the large $m\ell$ limit, the support of the integral in Eq. \eqref{eq:Sn_Ising} is localized near $\theta \simeq 0$, and a saddle-point analysis reveals that the two-particle contribution decreases exponentially, scaling as $\sim e^{-2m\ell}$. While these considerations hold for integer values of $n \geq 2$, the analytical continuation $n \rightarrow 1$ is more subtle, as we explain below.

We observe that $\frac{|F_{11}^{(n)}(\theta)|^2}{|\tau|^2}$ does not depend on the field normalisation explicitly, and it is fixed by bootstrap. Moreover, since for $n=1$ the twist field becomes the identity operator \cite{ccd-08}, its two-particle form factors vanish. This suggests that under analytic continuation over $n$, $F^{(n)}_{11}(\theta)$ converges to zero uniformly in the distributional sense (for $\theta$ real) as $n\rightarrow 1$ (see \cite{ccd-08}), a property that can be explicitly checked from the expression \eqref{two-kink form factor final}. Therefore, $\frac{|F^{(n)}_{11}(\theta)|^2}{|\tau|^2}$ behaves as $O((n-1)^2)$ for $(n-1)$ small, and, accordingly, the two-particle contribution to the von Neumann entropy, obtained as the limit $n\rightarrow 1$ in Eq. \eqref{eq:Sn_Ising}, vanishes identically. This circumstance is quite unexpected and, up to our knowledge, it has not been previously observed. Indeed, a two-particle contribution is always present in the paramagnetic phase of a large class of theories with a single vacuum, as shown in Ref. \cite{Doyon-09}. 

We finally stress that our analysis does not imply that the entanglement entropy is exactly independent of $\ell$. In fact, it is reasonable to expect a non-vanishing contribution from the four-kink form factors, thus yielding
\be
S_1 \simeq \text{const.} + O(e^{-4m\ell})\,.
\ee
However, the investigation of higher-kink form factors is beyond the purpose of this work.

\section{Form factors for the composite twist fields and entanglement asymmetry in the ferromagnetic phase of Ising}\label{sec:FF_comp_Ising}

In this section, we analyse the $\mathbb{Z}_2$ composite twist fields that arise in the computation of the entanglement asymmetry. These are the building blocks used to reconstruct the composite twist operators introduced in Sec. \ref{sec:twist_op}, and, as for the standard twist fields, the correspondence goes as
\be
\mathcal{T}_A^{\{g_1,\dots,g_n\}} \sim \mathcal{T}^{\{g_1,\dots,g_n\}}(x)\,, \quad A = (x,\,\infty)\,,
\ee
\be
\TT_A^{\{g_1,\dots,g_n\}} \sim \TT^{\{g_1,\dots,g_n\}}(0)\l\TT^{\{g_1,\dots,g_n\}}\r^\dagger (\ell)\,, \quad A = (0,\ell)\,,
\ee
and similarly for any union of disjoint intervals. Physically, $\mathcal{T}^{\{g_1,\dots,g_n\}}(x)$ introduces a branch cut along $(x,\infty)$, connecting the $j$th and $(j+1)$th replicas via the additional insertion of an Aharonov–Bohm flux $g_j$. In particular, to compute the entanglement asymmetry, we only need those composite twist fields with vanishing net flux (satisfying $g_n\cdots g_1 =1$). A fundamental observation pointed out in Sec. \ref{sec:twist_op} is that such composite twist fields can be related to the standard one via global unitary transformations induced by the symmetry. In particular, given the form factors of $\mathcal{T}(x)$, the ones of $\mathcal{T}^{\{g_1,\dots,g_n\}}(x)$ can be easily reconstructed once the action of the symmetry on the multi-particle states is known. Before entering the core of this section, we recall that $\mathbb{Z}_2$ composite twist fields with non-vanishing net flux have been characterised in the paramagnetic phase of the Ising model \cite{hc-20,cm-23} and used to compute the symmetry-resolved entanglement entropy: those fields are not related by global unitary transformations to $\mathcal{T}(x)$ and the discussion of this section does not apply to them. 

Our analysis refers to the ferromagnetic phase of the Ising field theory, where the group is $G= \mathbb{Z}_2$ which corresponds to a spin-flip exchanging the two vacua. We parametrise the elements of $\mathbb{Z}_2$ as
\be
\mathbb{Z}_2 = \{1,\mu\}\,,
\ee
with $\mu^2=1$, and in the forthcoming discussion, with a slight abuse of notation, we do not distinguish the elements of $\mathbb{Z}_2$ from the corresponding unitary operators (i.e. we drop the \lq\lq hat'' notation that we adopted in Sec. \ref{sec:twist_op}). It is easy to show that the $n$-tuples $\{g_j\}$ with vanishing net flux are the ones for which an even number of elements $\mu$ is present. Since $|G|=2$, there are $2^{n-1}$ of those $n$-tuples (see Eq. \eqref{Tr_rho_tilde_simplified}), and, for the sake of completeness, we list them explicitly in a few cases
\begin{itemize}
    \item $n=2$: $\{g_1, g_2\}$ can be $\{1,1\}$ and $\{\mu, \mu\}$.
    \item $n=3$: $\{g_1, g_2, g_3\}$ can be $\{1,1,1\}$, $\{1,\mu,\mu\}$ and its two cyclic permutations.
    \item $n=4$: $\{g_1, g_2, g_3, g_4\}$ can be $\{1,1,1,1\}$, $\{\mu,\mu,\mu,\mu\}$, $\{1,\mu,1,\mu\}$ and its cyclic permutation, $\{1,1,\mu,\mu\}$ and its three cyclic permutations.
\end{itemize}

We first discuss the zero-particle form factors of the composite twist fields. We anticipate that $\mathcal{T}^{\{g_1,\dots,g_n\}}(0)$ does not interpolate between ${}^{\otimes n}\bra{+}$ and any other vacuum (except in the trivial case $\{g_j=1\}$), a property which is crucial for the large-distance behavior of the entanglement asymmetry. Then, we move to the analysis of the two-particle form factors, providing exact analytical results. Finally, we collect these results together and give an expression for the R\'enyi entanglement asymmetry of a large interval.

\subsection{Zero-particle form factors}

We first recall the relation \eqref{unitary transformation simplified} and apply it to the composite twist fields, obtaining
\be\label{eq:un_transf_tfield}
\l g'_1 \otimes \dots \otimes g'_n \r \mathcal{T}(0) \l g'_1 \otimes \dots \otimes g'_n \r^{-1} = \mathcal{T}^{\{g_1,\dots,g_n\}}(0)\,, \quad g_j = g'_{j+1} (g'_j)^{-1}\,,
\ee
which is valid for $g_n\cdots g_1=1$. As previously established, the only non-vanishing vacuum expectation values of $\mathcal{T}(0)$ (see Eq. \eqref{TT vev}) are ${}^{\otimes n}\bra{\pm}\mathcal{T}(0)\ket{\pm}^{\otimes n}$. Similarly, thanks to \eqref{eq:un_transf_tfield}, the composite twist fields have only two non-vanishing zero-particle form factors: these are overlaps between the same vacuum, in general different from $\ket{\pm}^{\otimes n}$, as we show below. 

If $g'_j = 1$ or $g'_j=\mu$ for every $j=1,\dots,n$ in Eq. \eqref{eq:un_transf_tfield}, we have $\mathcal{T}^{\{g_1,\dots,g_n\}}(0) = \mathcal{T}(0)$. In all the other cases, namely if at least two elements of the $n$-tuple $\{g'_j\}$ are different, we get a non-trivial composite twist field $\mathcal{T}^{\{g_1,\dots,g_n\}}(x)$, i.e. at least one of the $g_j=\mu$. For these non-trivial cases, it holds
\be
{}^{\otimes}\bra{\pm} \mathcal{T}^{\{g_1,\dots,g_n\}}(0) \ket{\pm}^{\otimes n} = {}^{\otimes n}\bra{\pm} \l g'_1 \otimes \dots \otimes g'_n \r \mathcal{T}(0) \l g'_1 \otimes \dots \otimes g'_n \r^{-1} \ket{\pm}^{\otimes n} = 0,
\ee
since $\l g'_1 \otimes \dots \otimes g'_n \r^{-1} \ket{\pm}^{\otimes n}$ is always different from $\ket{\pm}^{\otimes n}$. This is one of the main results we will need later.

Other vacua have nevertheless non-vanishing amplitudes. Even if the latter do not appear in the computation of the entanglement asymmetry (as in Eq. \eqref{eq:asym_t_op} we take $\rho=\ket{+}\bra{+}$), it is useful to analyse a simple case explicitly. We consider for $n=2$ the twist field $\mathcal{T}^{\{\mu,\mu\}}(0)$. This is related to the standard twist field via the transformation \eqref{eq:un_transf_tfield}, with $\{g'_1 = 1, g'_2 = \mu\}$. Therefore, from \eqref{eq:un_transf_tfield} we get
\be
\bra{\pm, \mp} \mathcal{T}^{\{\mu,\mu\}}(0)\ket{\pm,\mp} = \bra{\pm,\pm}\mathcal{T}(0)\ket{\pm,\pm}\,,
\ee
and these are the only non-vanishing vacuum expectation values of $\mathcal{T}^{\{\mu,\mu\}}(x)$.

\begin{figure}[t]
    \centering
	\includegraphics[width=0.80\linewidth]{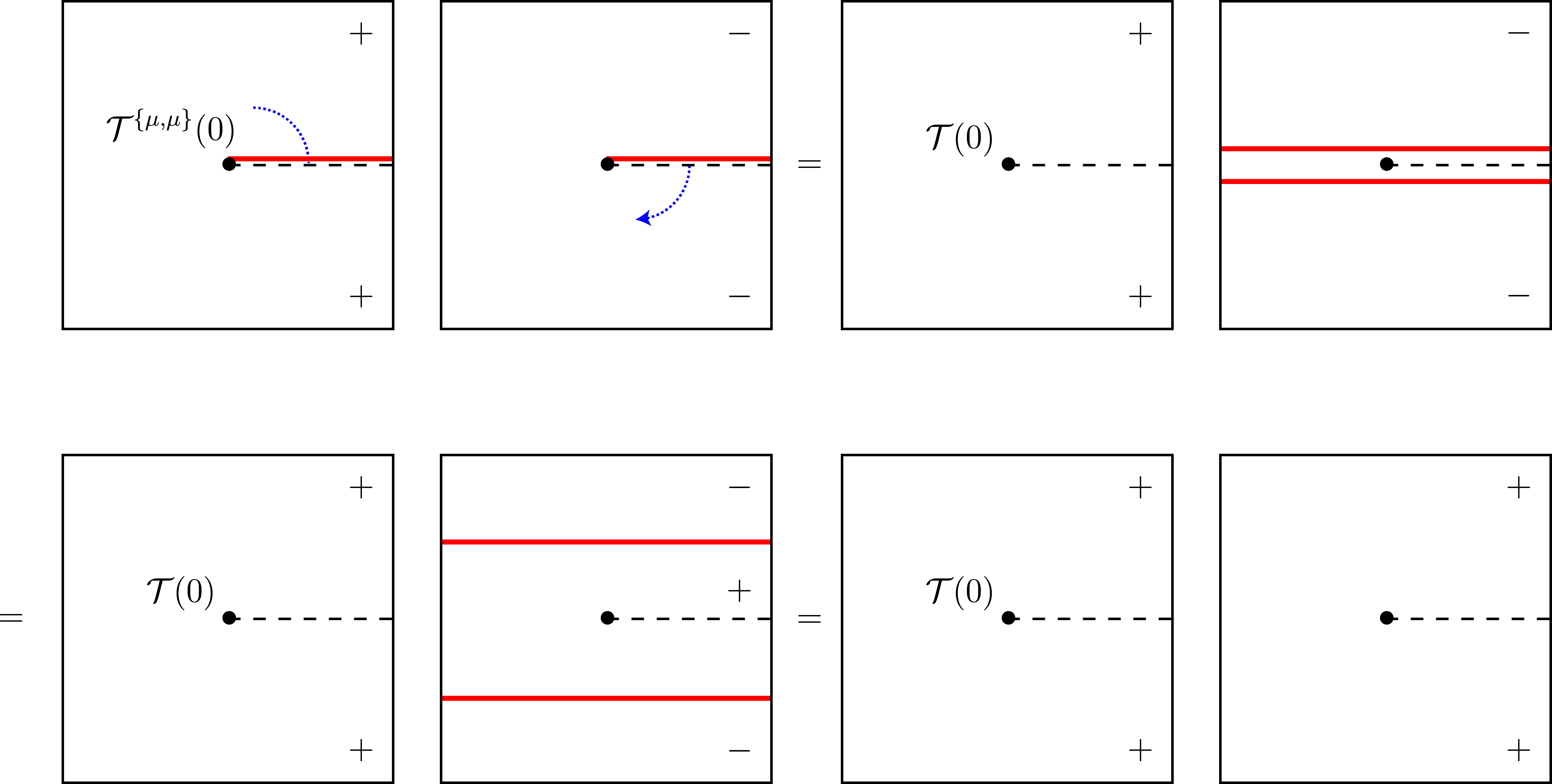}
    \caption{
    Pictorial representation of the form factor $\bra{+,-}\mathcal{T}^{\{\mu,\mu\}}(0)\ket{+,-}$. The blue dotted arrow in the top left figure shows that the second replica is reached from the first one when the branch cut, depicted by a dashed black line, is crossed clockwise. The red line represents the spin-flip, which exchanges $+$ and $-$. $\mathcal{T}^{\{\mu,\mu\}}(0)$ can be equivalently obtained via global spin-flips acting on $\mathcal{T}(0)$, as shown in the top right figure. Since the global spin-flip lines commute with the Hamiltonian, they can be translated and positioned at infinity, so that they act on the ingoing/outgoing states.
    This equivalence allows us to establish the relation $\bra{+,-}\mathcal{T}^{\{\mu,\mu\}}(0)\ket{+,-} = \bra{+,+}\mathcal{T}(0)\ket{+,+}$.
    }
    \label{fig:ff_Z2_tfield}
\end{figure}

We can interpret this result as follows. If the boundary conditions $+$ are chosen in the first replica, for both the outgoing and the incoming states, then, due to the spin-flip induced by the field, the boundary condition $-$ has to be present in the second replica. We represent this mechanism in Fig. \ref{fig:ff_Z2_tfield}, which shows pictorially the relation between $\mathcal{T}(x)$ and $\mathcal{T}^{\{\mu,\mu\}}(x)$.

\subsection{Two-particle form factors} \label{sec:2pff_composite}

The analysis for the two-particle form factors is slightly more involved. In principle, one can just use the form factors of Sec. \ref{sec:2p_ff}, with vacua as outgoing states and pairs of kinks as incoming ones, use the relation \eqref{eq:un_transf_tfield} and systematically reconstruct every non-vanishing two-particle form factor. In practice, this procedure becomes rather cumbersome as the number of composite twist operators grows exponentially with $n$.

For our purpose, it is more convenient to focus on the outgoing state ${}^{\otimes n}\bra{+}$ and the incoming states
\be
\ket{+,\dots,K_{+-}(\theta_1),\dots, K_{+-}(\theta_2),\dots,+},
\ee
with the kinks $K_{+-}(\theta_1)$, $K_{+-}(\theta_2)$ inserted in the replicas $j$ and $j'$ respectively and $+$ in every other replica. We do so, as these are the only contributions entering our computation of the entanglement asymmetry. Then, we investigate which $n$-tuples $\{g_j\}$ give rise to a non-vanishing form factor of $\mathcal{T}^{\{g_1,\dots,g_n\}}(x)$ for the states above. We also assume explicitly that $j\neq j'$ (say $j<j'$), since, for $j=j'$, that is when the two kinks are in the same replica, only the trivial twist field $\mathcal{T}(0)$ can interpolate between this state and  ${}^{\otimes n}\bra{+}$, a case already discussed in Sec. \ref{sec:2p_ff}.

\begin{figure}[t]
    \centering
	\includegraphics[width=0.80\linewidth]{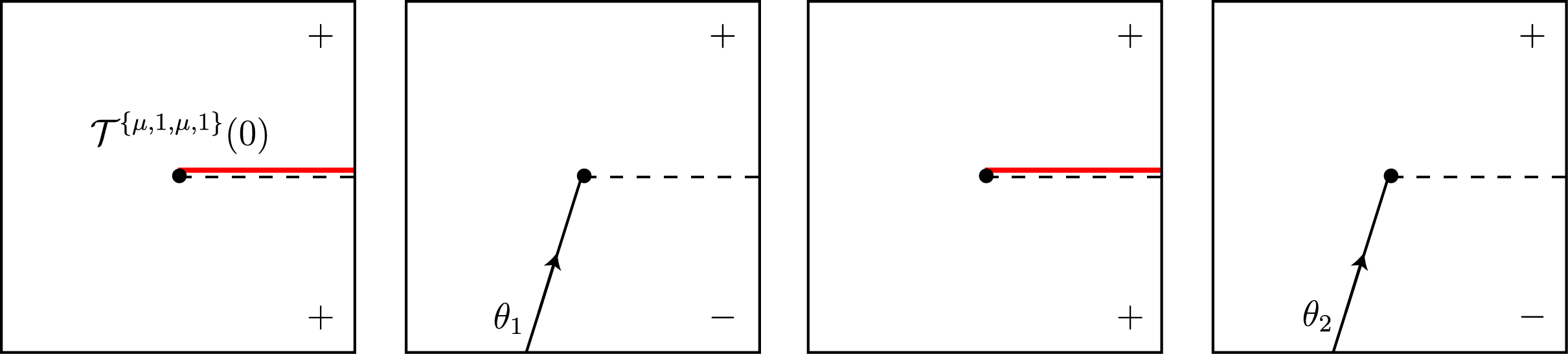}
    \caption{Two-particle form factor ${}^{\otimes 4}\bra{+}\mathcal{T}^{\{\mu,1,\mu,1\}}(0)\ket{+,K_{+-}(\theta_1),+,K_{+-}(\theta_2)}$. Since a single kink $K_{+-}$ is present in the second replica, a spin-flip, depicted as a red line, has to be inserted in the first replica  to get a non-vanishing form factor. Similarly, the presence of a kink in the fourth replica implies the spin-flip in the third replica.}
    \label{fig:ff_Z2_2pt}
\end{figure}

Let us first discuss the consequences of the presence of a (single) kink $K_{+-}$ in the $j$th replica. Since the boundary conditions are $\pm$ for $x= \mp \infty$, a spin-flip has to connect the $(j-1)$th and the $j$th replica. This is the only way to match the $+$ above the branch cut in the $(j-1)$th replica and the $-$ below the branch cut in the $j$th one. Vice versa, if no kinks are present in the $j$th replica, no spin-flips between the $(j-1)$th and the $j$th replicas can be present. An example of the mechanism above is shown in Fig. \ref{fig:ff_Z2_2pt}.

As a result, the only composite twist fields having non-vanishing form factors with two kinks in different replicas are precisely those with two spin-flips present. Namely, the matrix element
\be\label{eq:2p_ff_comp}
{}^{\otimes n}\bra{+} \TT(0)^{\{1,\dots,\mu,\dots, \mu,\dots,1\}} \ket{+,\dots, K_{+-}(\theta_1),\dots,K_{+-}(\theta_2),\dots,+},
\ee
with $\mu$ inserted in the $(j-1)$th, $(j'-1)$th replicas and $K_{+-}(\theta_1),\,K_{+-}(\theta_2)$ in the $j$th and $j'$th replica respectively, is non-vanishing. Furthermore, it is not difficult to show that the value of \eqref{eq:2p_ff_comp} is precisely $F^{(n)}_{11}(2\pi i(j'-j)-\theta_1+\theta_2)$, and we leave some details to Appendix \ref{app:two_p_FF}.

\subsection{Entanglement asymmetry}

In this section we put together all the results obtained so far, and we compute the entanglement asymmetry of an interval of length $\ell$ for the state $\ket{+}$. We focus on the limit $m\ell\gg 1$ keeping only the two-particle contribution. In this limit, the only twist fields appearing explicitly in the computation are the standard twist field $\mathcal{T}(x)$ and the composite fields with two spin-flips $\mathcal{T}^{\{1,\dots,\mu,\dots,\mu,\dots,1\}}(x)$. The reason is that the other fields do not have non-vanishing two-particle form factors with ${}^{\otimes n}\bra{+}$, and their contribution is therefore subleading.

For $A = (0,\ell)$ and $\rho=\ket{+}\bra{+}$ we can write
\begin{align}
 &\sum_{g_1,\dots,g_{n-1} \in \{1,\mu\}} \Tr\l\rho^{\otimes n} \TT_A^{\{g_1,\dots,g_{n-1},(g_1,\dots,g_{n-1})^{-1}\}}\r \nonumber \\=
&\sum_{g_1,\dots,g_{n-1} \in \{1,\mu\}} {}^{\otimes n}\bra{+} \TT_A^{\{g_1,\dots,g_{n-1},(g_1,\dots,g_{n-1})^{-1}\}}\ket{+}^{\otimes n} \\
\simeq &{}^{\otimes n}\bra{+}\mathcal{T}(0)\mathcal{T}^\dagger(\ell)\ket{+}^{\otimes n} + \sum_{1\leq j<j'\leq n } {}^{\otimes n}\bra{+}\mathcal{T}^{\{1,\dots,\mu,\dots,\mu,\dots,1\}}(0)\l\mathcal{T}^{\{1,\dots,\mu,\dots,\mu,\dots,1\}}\r^\dagger(\ell)\ket{+}^{\otimes n}\nonumber\,,
\end{align}
where the spin-flips $\mu$ are inserted at positions $j$ and $j'$. In the large volume limit $m\ell\rightarrow \infty$ the two-point function of the standard twist fields converges to $|\tau|^2$ (the square modulus of the vacuum expectation value). In contrast, the one of composite twist fields goes to zero exponentially fast as $\sim e^{-2m\ell}$, because these fields have vanishing vacuum expectation value and the first non-trivial contribution comes from the two-particle form factors. Expanding in the basis of the multi-kink states, we arrive at
\begin{align}
&\sum_{g_1,\dots,g_{n-1} \in \{1,\mu\}} \frac{{}^{\otimes n}\bra{+} \TT_A^{\{g_1,\dots,g_{n-1},(g_1,\dots,g_{n-1})^{-1}\}}\ket{+}^{\otimes n}}{ {}^{\otimes n}\bra{+} \TT_A\ket{+}^{\otimes n} }  \\
&\simeq 1+ \frac{n}{4\pi^2}\sum_{j=1}^{n-1} \int_{\mathbb{R}} \mathrm{d}\theta\,K_0\l 2 m \ell \cosh\l\frac{\theta}{2}\r\r \frac{|F_{11}^{(n)}(2\pi i j-\theta)|^2}{|\tau|^2}.
\end{align}
Finally, we employ the definition of R\'enyi asymmetry entanglement, and from \eqref{eq:asym_t_op} we get
\be\label{eq:DeltaSn_Ising}
\Delta S_n \simeq \log 2 - \frac{1}{n-1}\log \l  1+ \frac{n}{4\pi^2}\sum_{j=1}^{n-1} \int_{\mathbb{R}} \mathrm{d}\theta\,K_0\l 2 m \ell \cosh\l\frac{\theta}{2}\r\r \frac{|F_{11}^{(n)}(2\pi i j-\theta)|^2}{|\tau|^2} \r, 
\ee
an expression valid for any integer $n\geq 2$.

Before discussing the analytic continuation over $n$, some comments are in order. First, the R\'enyi entanglement asymmetry is a universal quantity as it does not depend on the normalisation of the twist fields: this is manifest in the appearance of the two-particle form factor $F^{(n)}_{11}(\theta)$ through its ratio with the VEV only. Moreover, for large $m\ell$, the quantity $\Delta S_n$ approaches its asymptotic value $\log 2$ from below, and this limit is compatible with the inequality
\be
\label{eq:asymmetry_bounds_main}
0 \leq \Delta S_n\leq \log |G|,
\ee
which we expect to have general validity, and we prove in some simple cases in Appendix \ref{app:inequalities}. These conclusions appear to be quite general, as they do not depend on the explicit  expression of the two-particle form factors in Eq. \eqref{two-kink form factor final}: instead, they strongly rely on the vanishing of the composite twist fields vacuum expectation values, a property stemming from topological constraints only.

In order to obtain the entanglement asymmetry we need to take the limit $n \to 1^+$ of $\Delta S_n$, which requires to analytically continue the sum
\be
\sum_{j=1}^{n-1} |F^{(n)}_{11}(2i\pi j- \theta)|^2,
\ee
a task that has been accomplished explicitly in \cite{ccd-08}. Following the same steps and employing the notation thereof, we define the function 
\be\label{eq:f_theta_n}
f(\theta, n)= \frac{1}{|\tau|^2}\sum_{j=0}^{n-1}|F_{11}^{(n)}(2\pi i j-\theta)|^2,
\ee
and we denote with $\tilde{f}(\theta, n)$ its analytic continuation from $n=2,3,\dots,$ to $n\in [1,\infty)$. In \cite{ccd-08} it has been shown that
\be
\lim_{n\to 1^+} \tilde{f}(\theta, n)=0\,, \quad \lim_{n\to 1^+} \frac{\partial\tilde{f}(\theta, n)}{\partial n} = \frac{\pi^2 }{2}\delta(\theta),
\ee
a relation valid in the distributional sense over real values of $\theta$. Here, we are interested in the sum
\be
g(\theta,n) = \frac{1}{|\tau|^2}\sum_{j=1}^{n-1}|F_{11}^{(n)}(2\pi i j-\theta)|^2 = f(\theta,n)-\frac{|F^{(n)}_{11}(\theta)|^2}{|\tau|^2},
\ee
namely, the term with $j=0$ in Eq. \eqref{eq:f_theta_n} is not present in our calculation. Since, as we showed in Sec. \ref{sec:entropy_analytic_continuaiton}, $|F^{(n)}_{11}(\theta)| \sim O((n-1)^2)$, the same properties established above for $\tilde{f}(\theta,n)$ are valid for $\tilde{g}(\theta,n)$, including its analytic continuation. In particular, we have\footnote{We mention that an explicit expression for $\tilde{g}(\theta,n)$ can be given for any $n \in [1,\infty)$. In particular, using the results of \cite{ccd-08} one has $\tilde{g}(\theta, n) 
= \tanh\l\frac{\theta}{2}\r \Im \l \frac{F_{11}^{(n)}(-2\theta+i\pi)-F_{11}^{(n)}(-2\theta+i\pi(2n-1))}{\tau}\r - \frac{|F_{11}^{(n)}(\theta)|^2}{|\tau|^2}$.}
\be
\lim_{n\to 1^+} \tilde{g}(\theta, n)=0\,, \quad \lim_{n\to 1^+} \frac{\partial\tilde{g}(\theta, n)}{\partial n} = \frac{\pi^2 }{2}\delta(\theta)\,.
\ee
Inserting this result in \eqref{eq:DeltaSn_Ising} we finally get\footnote{We recall the asymptotics of the Bessel function (see Ref. \cite{gr-14}) that is $K_0(z) \sim \sqrt{\frac{\pi}{2z}}e^{-z}$ for large $z$. In particular, the two-particle contribution to the entanglement asymmetry is exponentially suppressed in the size of the region.}
\begin{align}
\Delta S_1 &=  \log 2 -\frac{1}{4\pi^2} \lim_{n\to 1^+} \int_{\mathbb{R}} \mathrm{d}\theta\,K_0\l 2 m \ell \cosh\l\frac{\theta}{2}\r\r \frac{\partial}{\partial n} (n\tilde{g}(\theta,n)) + O(e^{-4m\ell}) \nonumber \\
&= \log 2 - \frac{K_0(2m\ell)}{8} + \mathcal{O}(e^{-4m\ell}),
\end{align}
that is the main result of this section.

\section{Further generalisations}\label{sec:Further_gen}

In this section, we propose a general conjecture regarding the entanglement asymmetry for any finite group $G$. We believe that our finding $\Delta S_n \simeq \log 2$ is not specific to the integrability features of the Ising field theory, and it mostly relies on the symmetry breaking pattern $\mathbb{Z}_2 \rightarrow \{1\}$ of this model. We also provide a paradigmatic case in which our conjecture can be proven with elementary techniques, that is, the case of zero-entanglement states.

Let us first recall that a state $\rho$ is symmetric under $g \in G$ if the following equality holds
\be
\rho = g\rho g^{-1}.
\ee
Clearly, the set of $g\in G$ leaving $\rho$ invariant is a group itself. Therefore, we define
\be
H \equiv  \{ h \in G | \rho = h\rho h^{-1}\} \subset G,
\ee
and we say that the symmetry-breaking pattern
\be
G\rightarrow H,
\ee
arises for the state $\rho$. We conjecture that the R\'enyi asymmetry of a large subsystem $A$ is
\be\label{eq:conj}
\Delta S_n \simeq \log \frac{|G|}{|H|}.
\ee
In particular, for the Ising model $G = \mathbb{Z}_2$, $H = \{1\}$ (the trivial group), and our conjecture is compatible with the main result \eqref{eq:Asym_Ising} of this paper. In general, $|G|/|H|$ is precisely the number of ground states (vacua) in the ordered phase of a theory. For example, for the $q$-state Potts model \cite{dc-98}, $q$ distinct vacua are present and our conjecture \eqref{eq:conj} gives  $\Delta S_n \simeq \log q$. We remark that the conjecture Eq. \eqref{eq:conj} refers to finite groups only, and it does not apply to the case of $U(1)$ that has been previously studied in Ref. \cite{amc-23} (see also Appendix \ref{app:inequalities}).

While we do not have a rigorous general proof of \eqref{eq:conj}, we think that a few physical hypotheses should be sufficient to prove it:
\begin{itemize}
\item Homogeneity: a finite region $A$ has to encode the global symmetries of the total system, that is the case for homogeneous systems.
\item Short-range correlations: we require that correlations among points of the region $A$ decay fast enough. This is the case for ground-states, and equilibrium states of short-range Hamiltonians. We believe that this hypothesis is sufficient to prove the existence of a large-volume limit for the entanglement asymmetry.
\end{itemize}

Before discussing an explicit case where the conjecture \eqref{eq:conj} can be proven, we would like to point out the intuition behind it. Let us first express, for $n\geq 2$
\be\label{eq:ratio_mom}
\frac{\text{Tr}(\tilde{\rho}^n_A)}{\text{Tr}(\rho^n_A)} = \frac{1}{|G|^{n-1}} \sum_{g_1,\dots,g_{n-1} \in G}\frac{\text{Tr}(\rho_A g_{1}\rho_A \dots g_{n-1}\rho_A (g_1\dots g_{n-1})^{-1})}{\text{Tr}(\rho_A^n)},
\ee
where both the index $A$ and the hat have been dropped from the operators for notational convenience. Whenever $g_j \in H$  in the sum above, that is
\be
\rho g_j = g_j \rho, \quad j=1,\dots,n-1,
\ee
one can commute the elements inside the trace in the numerator, obtaining
\be\label{eq:ratio_sym_elem}
\frac{\text{Tr}(\rho_A g_{1}\rho_A \dots g_{n-1}\rho_A (g_1\dots g_{n-1})^{-1})}{\text{Tr}(\rho_A^n)}=1, \quad g_j \in H.
\ee
There are precisely $|H|^{n-1}$ ways one can choose such $(n-1)$-tuples satisfying $\{g_j \in H\}_{j=1,\dots,n-1}$. The other terms in the sum \eqref{eq:ratio_mom} are expected to vanish in the large volume limit. Indeed, if $[\rho_A,g_j]\neq 0$ the eigenspaces of $g_j$ and $\rho_A$ are distinct: in the large volume limit, we expect that the eigenspaces above are at \lq\lq generic positions'' with respect to each other at large size, leading to a fast decay of the traces in which $\rho_A$ and $g_j$ are inserted consecutively: a related mechanism is known and established in the context of \textit{free independence} between random matrices, and we refer the interested reader to \cite{Voiculescu-06} for a review. The considerations above lead to \eqref{eq:conj} after simple algebra .

\subsection{Zero-entanglement states}

For zero-entanglement states, the conjecture \eqref{eq:conj} can be proven for any finite group $G$, as we show below. While this example may not be a realistic description of physical states with short-range correlations, it correctly captures the important features we mentioned above.

We start with the pure state
\be
\ket{0\dots 0}_A \otimes \ket{0 \dots 0}_{\bar{A}},
\ee
in which $\ket{0}$ is an on-site configuration belonging to a finite-dimensional Hilbert space. The reduced density matrix of the state above is
\be
\rho_A = \ket{0\dots 0}_A \prescript{}{A}{\bra{0\dots 0}},
\ee 
and, since $\rho_A$ is pure, no entanglement between $A$ and $\bar{A}$ is present. In particular, this means $\text{Tr}\l \rho^n_A\r=1$.

Let us assume that $G$ acts unitarily on the system above as a global symmetry. One can easily show that the group $H$ leaving the state is invariant is simply characterised as
\be
H = \{h \in G | \ |\bra{0}h\ket{0}|=1 \},
\ee
that is $\ket{0}$ and $h\ket{0}$ are proportional through a phase. In contrast, whenever $g\notin H$, one has
\be
|\bra{0}g\ket{0}|<1.
\ee 

We consider a generic $n$-tuple $\{g_j \in G\}_{j=1,\dots,n}$, and we compute
\be
\text{Tr}\l \rho_A g_1\dots \rho_A g_n\r = \prod^{n}_{j=1} {}_{A}\bra{0\dots 0}g_j\ket{0\dots 0}_A = \prod^{n}_{j=1} \bra{0}g_j\ket{0}^{|A|},
\ee
with $|A|$ the number of sites of $A$. As long as at least one element satisfies $g_j \notin H$, one has 
\be
\underset{|A|\rightarrow \infty}{\lim} |\bra{0}g_j\ket{0}|^{|A|} = 0,
\ee
which trivially implies
\be
\underset{|A|\rightarrow \infty}{\lim}  \text{Tr}\l \rho_A g_1\dots \rho_A g_n\r = 0.
\ee
On the other hand, for $n$-tuples of elements of $H$, Eq. \eqref{eq:ratio_sym_elem} holds. Putting together these results with \eqref{eq:ratio_mom}, one finally gets \eqref{eq:conj} for the case of a large subsystem $A$.

\section{Conclusions}\label{sec:conclusions}

In this paper we generalised the notion of R\'enyi entanglement asymmetry, first proposed in \cite{amc-23}, to include the action of an arbitrary finite group $G$, and we characterised its value in the symmetry-broken ground state of the Ising field theory. In particular, we employed the replica trick to describe the quantities of interest as expectation values of composite twist operators, and we provided analytical expressions of their form factors using integrable bootstrap. 

In addition, we proposed a general conjecture \eqref{eq:conj}, expected to hold for a large class of quantum states and for any finite group $G$. Remarkably, if our conjecture is correct, the entanglement asymmetry of a large but finite region becomes a useful tool to \lq\lq count'' the number of ground states in a spontaneously symmetry-broken phase via a local probe.

Many interesting directions are left open. It would be interesting to investigate other integrable QFTs using the approach we developed in this paper. For instance, the $q$-state Potts model in its ferromagnetic phase \cite{dc-98} stands out as the simplest generalisation of the Ising model. Furthermore, one could analyse massless flows \cite{dms-95} in which the symmetry is broken explicitly along the renormalisation group flow, and distinct symmetries are present in the IR and UV fixed points. It may also be worth considering field theories where the symmetry breaking arises due to the presence of boundary conditions or impurities (see e.g. Refs. \cite{gz-94,bcp-23}) and translational invariance is absent.

An intriguing puzzle is the relation between the paramagnetic and ferromagnetic phases at the level of entanglement. As we have shown, formal analogies are present in the form factor bootstrap of the twist fields, but the relation is subtle and, in general, the entropies are different in the two phases (see Appendix \ref{app:CTM}). As well known, Kramers-Wannier duality \cite{kw-41}, which is nowadays understood in the language of non-invertible defects \cite{amf-16}, relates those phases non-locally: this duality might be a good candidate to explain the relationship between the twist fields in the two phases. In particular, it would be interesting to understand what is the \lq\lq dual'' of the entanglement of a region under the Kramers-Wannier duality.

We also consider the study of low-energy states in the ferromagnetic phase of one-dimensional quantum systems (i.e. domain walls interpolating between distinct vacua) to be promising. It is not clear whether the results found in \cite{cdds-18a,cdds-18b,cdds-19a,cdds-19b} for quasi-particles hold eventually for these topological excitations.

We left open the proof of \eqref{eq:conj}, and a possible generalisation for continuous groups. For example, one could investigate its validity for Gibbs ensembles, Matrix Product States (MPS) \cite{Schollwock-11}, regularised boundary states of CFT \cite{Cardy-16}. Moreover, it would be intriguing to understand whether Eq. \eqref{eq:conj} is also valid for the steady states arising in the long-time dynamics, which is the main motivation behind the original formulation of the entanglement asymmetry \cite{amc-23}.

 \medskip {\bf Acknowledgements:} The authors are grateful to Pasquale Calabrese, David Horvath, Alessandro Santini, Guido Giachetti, Olalla Castro Alvaredo and Filiberto Ares for the useful discussions.
Michele Mazzoni is grateful for funding under the EPSRC Mathematical Sciences Doctoral Training Partnership EP/W524104/1 and wants to thank SISSA for hospitality during a one-month visit in which most of this
work was completed.
LC acknowledges support from ERC under Consolidator grant number 771536 (NEMO).
\noindent

\begin{appendix}
\section{Two-particle form factors of composite twist fields}\label{app:two_p_FF}

In this appendix, we relate explicitly the two-particle form factor of the composite twist fields and the ones of the standard twist field. In particular, we show in some specific cases the validity of the relation
\be
{}^{\otimes n}\bra{+} \TT(0)^{\{1,\dots,\mu,\dots, \mu,\dots,1\}} \ket{+,\dots, K_{+-}(\theta_1),\dots,K_{+-}(\theta_2),\dots,+} = F^{(n)}_{11}(2\pi i(j'-j)-\theta_1+\theta_2),
\ee
which was presented at the end of Sec. \ref{sec:2pff_composite}. A proof of the equation above for generic values of $n$, $j$, $j^\prime$ is easy to obtain by making use of the monodromy equation, global spin-flip invariance and the algebra \eqref{FZ_algebra}.

Let us consider $n=2,3,4$, and we focus on (see Sec. \ref{sec:twist_op})
\be
\mathcal{T}^{\{\mu,\mu\}}(x) = (1\otimes \mu)\mathcal{T}(x)(1\otimes \mu),
\ee
\be
\TT^{\{1,\mu,\mu\}}(x)= (1\otimes 1\otimes \mu) \TT(x) (1\otimes 1\otimes \mu),
\ee
\be
\TT^{\{\mu,1,\mu,1\}}(x)=(1\otimes\mu \otimes \mu \otimes 1) \TT(x) (1\otimes\mu \otimes \mu \otimes 1),
\ee
respectively.
For $n=2$ we have
\begin{align}
     &\bra{+,+} \TT^{\{\mu,\mu\}} (0)\ket{K_{+-}(\theta_1),K_{+-}(\theta_2)} \nonumber \\
     =&\bra{+,+}(1\otimes \mu) \TT(0) (1\otimes \mu)\ket{K_{+-}(\theta_1),K_{+-}(\theta_2)} \nonumber \\
     =&\bra{+,-} \TT(0) \ket{K_{+-}(\theta_1),K_{-+}(\theta_2)} \nonumber \\
     =&\bra{-,-} \TT(0) \ket{K_{-+}(\theta_2+2\pi i)K_{+-}(\theta_1),-} \nonumber \\
     =&\bra{+,+} \TT(0) \ket{K_{+-}(\theta_2+2\pi i)K_{-+}(\theta_1),+} = F_{11}^{(2)}(2\pi i +\theta_2-\theta_1)\,.
    \end{align}
In the previous steps we employed the monodromy equation for $\mathcal{T}(x)$ (discussed in Sec. \ref{sec:FF_Ising}) in going from the third to the fourth line, while we applied an overall spin-flip transformation to each replica to obtain the last line. Similar calculations are shown below for $n=3$:
\begin{align}
     &\bra{+,+,+} \TT^{\{1,\mu,\mu\}}(0) \ket{K_{+-}(\theta_2),+,K_{+-}(\theta_1)} \nonumber \\
     =&\bra{+,+,+}(1\otimes 1 \otimes \mu) \TT(0) (1\otimes 1 \otimes \mu)\ket{K_{+-}(\theta_2),+,K_{+-}(\theta_1)} \nonumber \\
     =&\bra{+,+,-} \TT(0) \ket{K_{+-}(\theta_2),+,K_{-+}(\theta_1)} \nonumber \\
     =&\bra{-,-,-} \TT(0) \ket{K_{-+}(\theta_1+4\pi i)K_{+-}(\theta_1),-,-} \nonumber \\
     =&\bra{+,+,+} \TT(0) \ket{K_{+-}(\theta_1+4\pi i)K_{-+}(\theta_2),+,+} \nonumber \\=& F_{11}^{(3)}(4\pi i +\theta_2-\theta) = F_{11}^{(3)}(2\pi i -\theta_1+\theta_2).
    \end{align}
Finally, for $n=4$ one has
\begin{align}
     &\bra{+,+,+,+} \TT^{\{\mu,1,\mu,1\}}(0) \ket{+,K_{+-}(\theta_1),+,K_{+-}(\theta_2)} \nonumber \\
     =&\bra{+,+,+,+}(1\otimes \mu\otimes \mu \otimes 1) \TT(0) (1\otimes \mu\otimes \mu \otimes 1)\ket{+,K_{+-}(\theta_1),+,K_{+-}(\theta_2)} \nonumber \\
     =&\bra{+,-,-,+} \TT(0) \ket{+,K_{-+}(\theta_1),-,K_{+-}(\theta_2)} \nonumber \\
     =&\bra{+,+,+,+} \TT(0) \ket{+,K_{+-}(\theta_2+4\pi i)K_{-+}(\theta_1),+,+} \nonumber \\
     =& F_{11}^{(4)}(4\pi i -\theta_1+\theta_2).
    \end{align}

\section{Useful inequalities}\label{app:inequalities}

In this appendix, we provide some upper and lower bounds for the R\'enyi entanglement asymmetry of finite groups. In particular, we focus on $n=2$, leaving some considerations for the other values of $n$ at the end of this section. The main result, proved below, is
\be\label{eq:Asym_2_app}
0\leq\Delta S_2\leq \log |G|,
\ee
with $\Delta S_2 =0$ iff $\rho_A = \tilde{\rho}_A$.

We first express $\Delta S_2$, defined in \eqref{eq:Ent_asymm}, as (see also Eq. \eqref{eq:asym_t_op})
\be\label{eq:Asym_2}
\Delta S_2 = \log |G| - \log \l \sum_{g \in G} \frac{\text{Tr}\l \rho_A g \rho_A g^{-1}\r}{\text{Tr}\l \rho^2_A\r}\r.
\ee
Therefore, to bound $\Delta S_2$ it is sufficient to analyse the possible values taken by
\be\label{eq:ratio_g}
\frac{\text{Tr}\l \rho_A g\rho_A g^{-1}\r}{\text{Tr}\l \rho^2_A\r},
\ee
as a function of $g \in G$. We observe that, since $\rho_A$ and $g\rho_A g^{-1}$ are both positive semi-definite, one can easily show that\footnote{Given $A,B$ positive semi-definite matrices, one has $\text{Tr}\l AB\r = \text{Tr}\l (\sqrt{A}\sqrt{B})(\sqrt{A}\sqrt{B})^\dagger\r \geq 0$.}
\be
\frac{\text{Tr}\l \rho_A g\rho_A g^{-1}\r}{\text{Tr}\l \rho^2_A\r}\geq 0.
\ee
Since the ratio in Eq. \eqref{eq:ratio_g} is $1$ for $g=1$ and positive semi-definite for the other values of $g$, we immediately get $\Delta S_2 \leq \log |G|$ from Eq. \eqref{eq:Asym_2}.

To prove $\Delta S_2\geq0 $, we apply the von Neumann's trace inequality \cite{Mirsky-75} to $\rho_A$ and $g\rho_A g^{-1}$, and we get
\be
\frac{\text{Tr}\l \rho_A g\rho_A g^{-1}\r}{\text{Tr}\l \rho^2_A\r}\leq 1,
\ee
and the bound is saturated iff $\rho_A$ and $g\rho_A g^{-1}$ share the same eigenvectors (which means they have to be equal, as their eigenvalues are always the same). Inserting the result above in Eq. \eqref{eq:Asym_2}, we obtain
\be
\Delta S_2\geq 0,
\ee
where the equality holds iff $\rho_A = g\rho_A g^{-1}$ for any $g \in G$. Therefore, from the definition \eqref{eq:rho_tilde}, we have $\Delta S_2 = 0$ iff $\rho_A = \tilde{\rho}_A$, which proves the main result \eqref{eq:Asym_2_app} of this appendix.

It is worth considering whether comparable strategies can be utilized to establish the inequality $0\leq \Delta S_n \leq \log |G|$ for other values of $n$. The first issue we encounter is that
\be
\frac{\text{Tr}\l \rho_A \cdot g_1\rho_A g^{-1}_1 \cdot g_2\rho_A g^{-1}_2\dots g_{n-1} \rho_A g^{-1}_{n-1} \r}{\text{Tr}\l \rho^n_A \r}
\ee
is not necessarily real for $n\geq 3$. For example, the matrices
\be
M_1 = \begin{pmatrix}1 & 0 \\ 0 & 0\end{pmatrix}, \quad M_2 = \frac{1}{2}\begin{pmatrix}1 & 1 \\ 1 & 1\end{pmatrix}, \quad M_3 = \frac{1}{2}\begin{pmatrix}1 & i \\ -i & 1\end{pmatrix}
\ee
are positive-semidefinite, having $0$ and $1$ as eigenvalues, but $\text{Tr}\l M_1 M_2 M_3\r = \frac{1+i}{4}\notin \mathbb{R}$. This implies that the previous technique, employed for $n=2$ to show $\Delta S_2\leq \log |G|$, does not apply directly to $n \geq 3$.

On the other hand, a proof of $\Delta S_n \geq 0$ valid for any, possibly non-integer, $n\geq 1$ can be given. The key idea is that the symmetrized state $\tilde{\rho}_A$ is generically more mixed (less pure) than $\rho_A$, and, therefore, it has more entropy. More precisely, one can show from \eqref{eq:rho_tilde} that the superoperator $\Phi$, defined as
\be
\Phi(\rho_A) \equiv \tilde{\rho}_A,
\ee
is completely positive and trace-preserving (CPT). Then, we use the monotonicity of the sandwiched R\'enyi divergence \cite{mr-15}, valid for CPT maps, and we apply it directly to $\rho_A$ and the infinite temperature state $\frac{\mathbbm{1}_A}{\text{Tr}\l \mathbbm{1}_A \r}$. As an immediate consequence, we get
\be
\Delta S_n\geq 0, \quad n\geq 1.
\ee
However, using this approach, it is not clear whether $\Delta S_n =0$ implies $\rho_A = \tilde{\rho}_A$.

As a last remark, we point out that for a compact Lie group, the number of elements is infinite and asymmetry does not have an upper bound. For instance, in a simple example analysed in \cite{amc-23}, the $U(1)$ asymmetry grows logarithmically in the subsystem size. The exploration of compact Lie groups will is deferred to future works.

\section{Exact results for the R\'enyi entropies via corner transfer matrix and Kramers-Wannier duality}\label{app:CTM}

In this appendix, we review some exact results available for the entanglement of the Ising chain, following closely Ref. \cite{ccp-10}. Let us consider the quantum one-dimensional Ising model, described by the Hamiltonian
\be\label{eq:Is_Ham}
H = -\sum_j \sigma_j^x \sigma^x_{j+1} + h\sigma^z_j,
\ee
with $\sigma^x_j,\sigma^z_j$ the Pauli matrices at position $j$. This model displays a quantum critical point at $h = 1$ separating a ferromagnetic
phase ($h < 1$) from a paramagnetic one ($h > 1$). It is well known that Kramers-Wannier duality \cite{kw-41} relates the spectra of the Hamiltonian at $h$ and $h^{-1}$, and, more in general, local and semi-local observables are mapped into each other by the duality above. The same mechanism is observed in the underlying Ising field theory, where one can relate the paramagnetic and the ferromagnetic phase at the same value of the mass $m$. However, as Ref. \cite{ccp-10} shows, the entropies of dual points differ explicitly. In particular, the entanglement of the half-chain has been characterised analytically in the infinite-volume limit via the Corner Transfer Matrix. Here, we only report and discuss the final result found in \cite{ccp-10}.

Let us define the parameters
\be\label{eq:param}
k \equiv \text{min}(h,h^{-1}), \quad \epsilon = \pi \frac{K(\sqrt{1-k^2})}{K(k)},
\ee
with $K(z)$ the complete elliptic integral of the first kind. Then, given
\be
\epsilon_j \equiv \begin{cases} 
(2j+1)\epsilon \ &\quad h>1,\\
2j\epsilon \ &\quad h<1,
\end{cases}
\ee
for $j=0,1,\dots$ one can express the corresponding $n$th R\'enyi entropy as
\be\label{eq:Sn_Baxter}
S_n(h) = \frac{1}{1-n} \sum^{\infty}_{j=0} \log\frac{1+e^{-\epsilon_j n}}{(1+e^{-\epsilon_j})^n}.
\ee
Clearly, as shown in Eq. \eqref{eq:param}, the value of the parameter $\epsilon$ is the same at the dual points $h,h^{-1}$, even though the \lq\lq single-particle eigenvalues'' $\epsilon_j$ are different.

We emphasise that for $h<1$, corresponding to the ferromagnetic phase, the ground state is degenerate and one should be careful. For instance, the result \eqref{eq:Sn_Baxter} refers to the symmetric ground state (often dubbed GHZ state), which corresponds to
\be\label{eq:GHZ}
\ket{\text{GHZ}} = \frac{1}{\sqrt{2}}\l \ket{+}+\ket{-}\r
\ee
in our notation. To obtain the entropies of the symmetry broken ground state $\ket{\pm}$ one has to remove explicitly the zero-mode $j=0$ from the sum in \eqref{eq:Sn_Baxter}, as explained in \cite{ccp-10}, thus getting
\be\label{eq:Sn_SSB}
S_n(h<1) = \frac{1}{1-n} \sum^{\infty}_{j=1} \log\frac{1+e^{-\epsilon_j n}}{(1+e^{-\epsilon_j})^n} = \frac{1}{1-n} \sum^{\infty}_{j=0} \log\frac{1+e^{-\epsilon_j n}}{(1+e^{-\epsilon_j})^n} -\log 2,
\ee
which means that the R\'enyi entropy of the doubly degenerate state $\ket{\pm}$ is smaller than that of the GHZ state \eqref{eq:Sn_Baxter}, as one would intuitively expect.

Remarkably, in the limit $h\rightarrow 1$ one has $\epsilon\rightarrow 0$ and, accordingly, the R\'enyi entropy diverges. The origin of the divergence above is universal, and it is traced back to the properties of the underlying CFT (see \cite{ccp-10} for further details): in particular, the divergence is the same if the critical magnetic field $h=1$ is approached from above or below.

Discrepancies between dual points can be nevertheless spotted if one takes into account also the finite terms arising in \eqref{eq:Sn_Baxter} when taking the limit $h\rightarrow 1$. In particular, using the results of \cite{ccp-10} one can show that the entropy difference between the paramagnetic phase and the corresponding ferromagnetic one (in the GHZ state) satisfies
\be\label{eq:S_crit}
\underset{h \rightarrow 1^+} {\lim} S_n(h)-S_n(h^{-1}) = - \frac{\log(2)}{2}.
\ee
The origin of the constant $-\frac{\log(2)}{2}$ is much more subtle with respect to the $\log 2$ term in \eqref{eq:Sn_SSB}: indeed, the former appears only close to the critical point, and a full explanation based on the formal analogy between \eqref{eq:Sn_Baxter} and the thermal entropy of a CFT has been provided in \cite{ccp-10} (see also Ref. \cite{cd-09} for a related discussion).

Below, we interpret these results in the language of the Ising field theory. Let us focus on a given value of the mass $m$ and let $\ket{0}$ be the paramagnetic ground state, $\ket{\pm}$ the symmetry broken ferromagnetic ground state, and $\ket{\text{GHZ}}$ the state defined by \eqref{eq:GHZ}. To compute the R\'enyi entropy of a given state in the half-infinite chain, say $A = (0,\infty)$, we have to evaluate the twist field $\mathcal{T}(0)$ over the corresponding replica state. In particular, the difference of R\'enyi entropy between $\ket{\text{GHZ}}$ and $\ket{+}$ would be just
\be\label{eq:GHZ_log2}
\frac{1}{1-n}\log \frac{{}^{\otimes n}\bra{\text{GHZ}}\mathcal{T}(0)\ket{\text{GHZ}}^{\otimes n}}{{}^{\otimes n}\bra{+}\mathcal{T}(0)\ket{+}^{\otimes n}} = \log 2 .
\ee
Here, to obtain \eqref{eq:GHZ_log2} we only used the definition \eqref{eq:GHZ} and the fact that the only non-vanishing VEVs of $\mathcal{T}(0)$ over replica vacua are ${}^{\otimes n}\bra{+}\mathcal{T}(0)\ket{+}^{\otimes n} = {}^{\otimes n}\bra{-}\mathcal{T}(0)\ket{-}^{\otimes n}$ (see Sec. \ref{sec:FF_Ising}). This result is compatible with \eqref{eq:Sn_SSB}, and reasonably a similar argument can be employed even beyond the field theoretic regime, when for $|h-1|$ is finite. 

Instead, a different mechanism arises if we try to compare $\ket{+}$ and $\ket{0}$. We first notice that the ratio
\be
\frac{{}^{\otimes n}\bra{+}\mathcal{T}(0)\ket{+}^{\otimes n}}{{}^{\otimes n}\bra{0}\mathcal{T}(0)\ket{0}^{\otimes n}}
\ee
is, by dimensional analysis, dimensionless, i.e. a universal number which does not depend on $m$ (the only mass scale). Moreover, according to the lattice result \eqref{eq:S_crit} valid near the critical regime where the field theory is predictive, we expect
\be\label{eq:g_affleck}
\frac{{}^{\otimes n}\bra{+}\mathcal{T}(0)\ket{+}^{\otimes n}}{{}^{\otimes n}\bra{0}\mathcal{T}(0)\ket{0}^{\otimes n}} = g^{n-1}, \quad g\equiv\sqrt{2},
\ee
so that the difference of entropy between $\ket{+}$ and $\ket{0}$ would be just\footnote{We recall that \eqref{eq:S_crit} compares the GHZ state and the paramagnetic one. If we take $\ket{+}$, instead of the GHZ, we get $\underset{h \rightarrow 1^+} {\lim} S_n(h)-S_n(h^{-1}) =  \frac{\log(2)}{2}$.}
\be
\frac{1}{1-n}\log \frac{{}^{\otimes n}\bra{+}\mathcal{T}(0)\ket{+}^{\otimes n}}{{}^{\otimes n}\bra{0}\mathcal{T}(0)\ket{0}^{\otimes n}} = -\frac{\log 2}{2}.
\ee
We believe that \eqref{eq:g_affleck} can be proven, once the normalisation of the twist field is fixed, via a form factor approach similar to that of \cite{cd-09}. We also conjecture that the value of the constant $g$ in Eq. \eqref{eq:g_affleck} comes precisely from Affleck-Ludwig boundary entropy \cite{al-91}: however, it is not clear to us whether an explicit relation between the off-critical paramagnetic/ferromagnetic phases and a boundary CFT can be provided. We only point out that if \eqref{eq:g_affleck} is correct, then it cannot be a trivial consequence of the Kramers-Wannier duality, which holds exactly on the lattice, as the relation \eqref{eq:S_crit} is only valid close to the critical point.

\end{appendix}

\printbibliography

\end{document}